\documentclass[12pt]{article}
\pdfoutput=1 
\usepackage{amsmath}
\usepackage{graphicx,psfrag,epsf}
\usepackage{enumerate}
\usepackage{url} 
\usepackage{natbib}
\setcitestyle{aysep={}} 
\usepackage{graphicx,verbatim,array,multicol, palatino}
\usepackage{amsthm, amsmath, amssymb,  setspace,natbib}
\usepackage{epsfig, url,epstopdf}
\usepackage{indentfirst,enumerate}
\usepackage{color}
\usepackage[super]{cite}
\usepackage[toc]{appendix}
\newtheorem{theorem}{Theorem}
\usepackage[top    = 1.1in, bottom = 1.1in, left   = 1in, right  = 1in]{geometry}

\renewcommand{\baselinestretch}{1}


\usepackage{color}
\definecolor{red}{rgb}{1,0,0}
\definecolor{blue}{rgb}{0,0,1}
\definecolor{green}{rgb}{0,0.6,0.4}
\def\red{\color{red}}

\newcommand{\blind}{1}

\begin{document}
\pagenumbering{gobble} 

%%%%%%%%%%%%%%%%%%%%%%%%%%%%%%%%%%%%%%%%%%%%%%%%%%%%%%%%%%%
% JASA spacing do not change
\def\spacingset#1{\renewcommand{\baselinestretch}%
{#1}\small\normalsize} \spacingset{1}
%%%%%%%%%%%%%%%%%%%%%%%%%%%%%%%%%%%%%%%%%%%%%%%%%%%%%%%%%%%

%%%%%%%%%%%%%%%%%%%%%%%%%%%%%%%%%%%%%%%%%%%%%%%%%%%%%%%%%%%
% JASA blinding title page
\if1\blind
{
  \title{\bf Set-Based Tests for Genetic Association Using the Generalized Berk-Jones Statistic}
  \author{Ryan Sun and Xihong Lin
 \thanks{Ryan Sun is a postdoctoral fellow and Xihong Lin is Professor, both with Department of Biostatistics, Harvard T.H. Chan School of Public Health, Boston, MA 02115.  This work was supported by the National Institutes of Health: R35-CA197449, P01-CA134294, and U01-HG009088-0.}\hspace{.2cm}\\
 }
  \maketitle
} \fi

\if0\blind
{
  \bigskip
  \bigskip
  \bigskip
  \begin{center}
    {\LARGE\bf Set-Based Tests for Genetic Association Using the Generalized Berk-Jones Statistic}
\end{center}
  \medskip
} \fi
%%%%%%%%%%%%%%%%%%%%%%%%%%%%%%%%%%%%%%%%%%%%%%%%%%%%%%%%%%%

%%%%%%%%%%%%%%%%%%%%%%%%%%%%%%%%%%%%%%%%%%%%%%%%%%%%%%%%%%%
% JASA abstract page

\bigskip
\begin{abstract}
Studying the effects of groups of Single Nucleotide Polymorphisms (SNPs), as in a gene, 
genetic pathway, or network, can provide novel insight into complex diseases, above that 
which can be gleaned from studying SNPs individually. 
Common challenges in set-based genetic association testing include weak effect sizes,
correlation between SNPs in a SNP-set, and scarcity of signals, with single-SNP effects
often ranging from extremely sparse to moderately sparse in number.
Motivated by these challenges, we propose the Generalized Berk-Jones (GBJ) test for 
the association between a SNP-set and outcome. 
The GBJ extends the Berk-Jones (BJ) statistic by accounting for correlation among 
SNPs, and it provides advantages over the Generalized Higher Criticism (GHC) test 
when signals in a SNP-set are moderately sparse.  
We also provide an analytic p-value calculation procedure for SNP-sets of any finite size.
Using this p-value calculation, we illustrate that the rejection region for GBJ can be described
as a compromise of those for BJ and GHC.
We develop an omnibus statistic as well, and we show that this omnibus test is robust to 
the degree of signal sparsity.
An additional advantage of our method is the ability to conduct inference using individual 
SNP summary statistics from a genome-wide association study.
We evaluate the finite sample performance of the GBJ though simulation studies and 
application to gene-level association analysis of breast cancer risk.
\end{abstract}

\noindent%
\vfill

\newpage
\spacingset{1.45} % DON'T change the spacing!
%%%%%%%%%%%%%%%%%%%%%%%%%%%%%%%%%%%%%%%%%%%%%%%%%%%%%%%%%%%

\pagenumbering{arabic}
\baselineskip=12pt

%%%%%%%%%%%%%%%%%%%%%%%%%%%%%%%%%%%%%%%%%%%%%%%%%%%%%%%%%%%
% Paper begins here.
\section{Introduction}
\label{p2_sec:intro}

Genome-Wide Association Studies (GWAS) have been successful in identifying the 
associations between thousands of Single Nucleotide Polymorphisms (SNPs) and 
a variety of complex traits \citep{Manolio_missing_herit}.  
A traditional GWAS analysis tests for the effect of each individual SNP, and this 
approach has shown that single-SNP effects are often weak across the genome 
\citep{Visscher_five}.  
Recently, set-based tests which jointly analyze a group of SNPs - e.g. SNPs in a 
gene, pathway, or network - have become increasingly popular as complementary 
tools which can boost analysis power in GWAS \citep{snp_set}.
These tests are also standard for rare variant analysis in whole genome sequencing 
studies \citep{rare_variant_review}.

SNPs can be aggregated into sets based on a variety of genomic features.
For example, they can be grouped by physical position, such as location in a gene 
or Linkage Disequilibrium (LD) block, or similar biological functions, such as membership
in a genetic pathway or protein network.
Set-based analyses then allow for some natural advantages over individual SNP methods. 
Besides reducing the number of multiple comparisons across the genome, SNP-set 
methods can increase power by pooling sparse and weak effects into a stronger aggregated 
signal, as well as by incorporating biological information into the test \citep{SKAT}. 
In addition, set-based interpretations of association may be more meaningful than their 
single-marker counterparts, such as in a gene-level or pathway-level analysis.

A number of set-based tests for genetic association studies have been developed in 
recent years, including burden tests \citep{LiLeal}, the Sequence Kernel Association 
Test (SKAT) \citep{SKAT}, the minimum p-value test (MinP) \citep{minP}, and most 
recently the  Generalized Higher Criticism (GHC) \citep*{GHC}. 
SKAT and burden tests are examples of methods more suitable for detecting dense signals.
If the signals reside in only a few SNPs that are not correlated with noise SNPs, then
the power of SKAT and burden tests will suffer.

While certain SNP-sets may contain a large number of signals, it is more common that 
genomic constructs formed with GWAS data will have only a few signal SNPs.
Interestingly, within tests designed for this sparse alternative setting, there are still subtle differences 
in performance.  
Under extreme sparsity, as in the case of only one or very few signals in the entire set, the 
minimum p-value test and the GHC have good power for detecting a SNP-set effect. 
However GHC and MinP can lose power under moderate sparsity settings, which are 
relatively common in gene and pathway level analyses of GWAS data.
For example, in the Cancer Genetic Markers of Susceptibility (CGEMS) GWAS for 
breast cancer risk \citep{CGEMS}, four out of 42 SNPs in the FGFR2 gene, a known 
breast cancer risk loci, showed strong evidence of association without reaching 
genome-wide significance.
It is hence of substantial interest to develop testing procedures that can reliably detect 
associations across a range of alternatives in the sparse signal regime.

When factors in a set are independent, several goodness-of-fit type methods have been 
proposed to perform set-based tests in the presence of sparse signals \citep{HC, Phi_divergences, ALR}.
These methods test for the effect of a set by aggregating evidence from marginal test 
statistics, and they have been shown to possess attractive asymptotic properties when 
the size of a set goes to infinity.
Specifically, they reach a so-called detection boundary when signals are sparse.
In a certain sense, they are able to detect the weakest signals detectable by any 
statistical procedure under the sparse alternative.
The class of tests with this ability includes the Higher Criticism (HC) \citep{HC} and 
the Berk-Jones (BJ) \citep{BJ}.
Compared to the HC and BJ tests, the minimum p-value test, for example, is known 
to attain the detection boundary over a smaller portion of the sparse regime.
In terms of finite sample performance, it has been demonstrated through simulation 
that Berk-Jones outperforms Higher Criticism over a range of moderately sparse 
alternatives when marginal test statistics are independent \citep{ALR, LiSiegmund}.
\citet{HC} provide an explanation for this result by showing that HC disproportionately 
weights evidence from the most extreme observed marginal test statistic, at the cost 
of losing sensitivity to detect signals in other locations. 

However, a direct application of BJ to SNP-set testing is not desirable, as standard 
p-value calculations for BJ rely on independence between elements in a set \citep{boundary_crossing}. 
This assumption is violated by LD-induced correlation between neighboring SNPs 
in a SNP-set.
In addition, we will see that even if we correct the inference procedure of the Berk-Jones, 
the power of BJ can fall dramatically in high LD settings.

To overcome the challenges posed by correlated SNPs, this paper proposes the 
Generalized Berk-Jones (GBJ) statistic for testing the association between a SNP-set 
and outcome. 
GBJ accounts for LD among SNPs in a set while still preserving the ability to detect 
moderately sparse and weak signals in finite samples. 
In fact, GBJ reduces to the Berk-Jones statistic when all SNPs in a set are independent. 
GBJ can also be applied to SNP-set tests using both individual-level genotype data 
or GWAS summary statistics from single SNP analysis.
To facilitate use, we additionally provide an analytic p-value calculation for GBJ. 
Our method is more computationally efficient than permutation and is shown to 
be accurate even at the extremely small levels required for genome-wide significance. 

Additional insight into the strengths and weaknesses of GBJ is provided by studying 
the rejection regions of SNP-set tests developed for sparse alternatives.
The rejection regions allow us to quantitatively describe how the power of each test is 
susceptible to changes in parameters such as the  amount of correlation between 
SNPs or the size of the SNP-set.
Since in practice we never have knowledge of the type of alternative, we also 
propose an omnibus test that combines GBJ, GHC, MinP, and SKAT for added 
robustness to different degrees of sparsity.
An extensive simulation study demonstrates that GBJ outperforms alternative 
methods in testing SNP-set effects when signals are weak and moderately sparse, 
and we also show that the omnibus test is robust to a wide range of sparsity levels.

The remainder of the paper is organized as follows. 
Section \ref{p2_sec:framework} discusses the SNP-set testing framework using both individual-level data 
and GWAS summary statistics. 
In Section \ref{p2_sec:GBJ} we propose the Generalized Berk-Jones statistic for testing the association 
between a SNP-set and outcome.  
We also provide an analytic p-value calculation for GBJ and develop the omnibus test.
Section \ref{p2_sec:rej_region} compares the rejection regions of GBJ and other tests designed for the sparse regime. 
In Section \ref{p2_sec:simulation} we demonstrate the finite sample performance of GBJ through simulation.
Section \ref{p2_sec:CGEMS} presents an analysis of the CGEMS data, and we conclude with a 
discussion in Section \ref{p2_sec:discussion}.

%%%%%%%%%%%%%%%%%%%%%%%%%%%%%%%%%%%%%%%%%%%%%%%%%%%%%%%%%%%%
%%%%%%%%%%%%%%%%%%%%%%%%%%%%%%%%%%%%%%%%%%%%%%%%%%%%%%%%%%%%
\section{The SNP-Set Testing Framework}
\label{p2_sec:framework}

\subsection{Individual-Level Genotype Data}
\label{p2_ss:IL_framework}

We begin by describing the SNP-set testing framework using individual-level data on 
genotype, outcome, and other covariates for $n$ total subjects.
Suppose for subject $i$, $i=1,...,n$, we observe the outcome $Y_{i}$, a genotype vector 
$\mathbf{G}_{i}=(G_{i1},...,G_{id})^{T}$ of $d$ SNPs in a SNP-set, and a vector of $q$ 
covariates $\mathbf{X}_{i}=(X_{i1},...,X_{iq})^{T}$.
Assume that $Y_{i}$ conditional on $(\mathbf{G}_{i},\mathbf{X}_{i})$ follows a distribution 
in the exponential family \citep{GLM} with the density function 
$f(Y_{i})=\exp\left\{ (Y_{i}\theta_{i}-b(\theta_{i}))/a_{i}(\phi)+c(Y_{i},\phi)\right\} $, where  
$a(\cdot)$, $b(\cdot)$, and $c(\cdot)$, are known functions, $\theta_{i}$ is a canonical 
parameter, and $\phi$ is a dispersion parameter. 
Let $\mu_{i}=E(Y_{i}|\mathbf{G}_{i},\mathbf{X}_{i})$ denote the conditional mean of 
$Y_{i}$ and assume it follows the Generalized Linear Model (GLM)
\[
g(\mu_{i})=\mathbf{X}_{i}^{T}\boldsymbol{\alpha}+\mathbf{G}_{i}^{T}\boldsymbol{\beta},
\]
where $g(\cdot$) is a differentiable monotone link function. 
We here only consider canonical link functions for simplicity. In matrix notation, the  
data take the form $\mathbf{Y}=(Y_{1},...,Y_{n})^{T}$,
$\mathbf{G}_{n\times d}=[\begin{array}{ccccc} \mathbf{G}_{1},..., \mathbf{G}_{n}\end{array}]^{T}$, 
and $\mathbf{X}_{n\times q}=[\begin{array}{ccccc} \mathbf{X}_{1},..., \mathbf{X}_{n}\end{array}]^{T}.$

The null hypothesis of no association between a SNP-set and outcome, after controlling 
for covariates, is given by $H_{0}:\boldsymbol{\beta}=\mathbf{0}$.  
Both the size of the set $d$ and the sparsity of signals can vary greatly between sets, 
e.g. from gene to gene, and the number of nonzero $\beta_{j}$ is unknown.
Our aim is to develop a test suitable for different levels of sparsity while also accounting 
for the correlation among individual SNP test statistics.

The marginal score statistic for $\beta_{j}$, $j=1,...,d$, is 
\[
Z_{j}=\frac{\mathbf{G}_{.j}^{T}(\mathbf{Y}-\widehat{\boldsymbol{\mu}}_{0})}{\sqrt{\mathbf{G}_{.j}^{T}\mathbf{P}\mathbf{G}_{.j}}},
\]
where $\mathbf{G}_{.j}$ denotes the $j$th column vector of $\mathbf{G}$, 
$\mathbf{P}=\mathbf{W}-\mathbf{W}\mathbf{X}(\mathbf{X}^{T}\mathbf{W}\mathbf{X})^{-1}\mathbf{X}^{T}\mathbf{W}$ 
is the projection matrix,
$\mathbf{W}=\text{diag}\left\{ a_{1}(\widehat{\phi}) v(\widehat{\mu}_{01}),...,a_{n}(\widehat{\phi}) v(\widehat{\mu}_{0n})\right\}$,
$\widehat{\mu}_{0i}=g(\mathbf{X}_{i}^{T}\widehat{\boldsymbol{\alpha}}_{0})$,
$\widehat{\boldsymbol{\alpha}}_{0}$ is the MLE of $\boldsymbol{\alpha}$
under the null hypothesis, and $v(\mu_{i})=b''(\theta_{i})$
is the variance function.

Under the null, the $d$ marginal score test statistics have an asymptotic multivariate normal distribution
\begin{eqnarray*}
\mathbf{Z}=\left( Z_{1},\cdots, Z_{d} \right)^{T} \overset{H_{0}}{\sim} N(\mathbf{0}_{d\times1},\boldsymbol{\Sigma}_{d\times d}),
\label{p2_eq:Z_dist}
\end{eqnarray*}
where $\Sigma_{jj}=1$ for all $j$, and for $j\neq k$ we can estimate
\begin{equation}
\widehat\Sigma_{jk}=\frac{\mathbf{G}_{.j}^{T}\mathbf{P}\mathbf{G}_{.k}}{\sqrt{\mathbf{G}_{.j}^{T}\mathbf{P}\mathbf{G}_{.j}}\sqrt{\mathbf{G}_{.k}^{T}\mathbf{P}\mathbf{G}_{.k}}}.
\label{p2_eq:estimate_cov}
\end{equation}

\subsection{GWAS Summary Statistics}
\label{p2_ss:summary_stats}

Many GWAS may not release individual-level data due to logistical challenges or data 
confidentiality agreements.
Instead it is much more likely that a marginal test statistic for association with the 
outcome is available for each individual SNP \citep{summary_statistic_review}.
It is hence of great interest to be able to perform SNP-set testing using precomputed
$Z_j$ from across the genome.
To test a set of precomputed $Z_{j}$ with GBJ, we require estimation of their 
correlation matrix $\boldsymbol{\Sigma}$ using external information.

Assume we have a panel of reference genotypes from $n_{r}$ subjects of the same 
ethnicity as those used to construct the summary statistics. 
For example, this could come from the publicly available 1000 Genomes dataset 
\citep{1000_Genomes}.  
We estimate $\boldsymbol{\Sigma}$ using equation (\ref{p2_eq:estimate_cov}) but replace  
$\mathbf{G}_{.j}$ and $\mathbf{X}$ with $\mathbf{G}_{.j}^{(r)}$  and $\mathbf{X}^{(r)}$, 
where $\mathbf{G}_{.j}^{(r)}$ is the $n_{r}\times 1$ genotype vector of SNP $j$ from 
the reference panel,   
$\mathbf{X}^{(r)}=\left(\mathbf{1}, \mathbf{PC}_{1},..., \mathbf{PC}_{m}\right)$, $\mathbf{PC}_{1},...,\mathbf{PC}_{m}$
are the first $m$ principal component vectors calculated from the reference panel, and 
$m$ is the same number of principal components as was used to control for population 
stratification \citep{Eigenstrat} in the original GWAS analysis of the data.  
Additionally for each subject we estimate $v(\widehat\mu_{0i})$ by setting $\widehat{\mu}_{0i}$ 
equal to the sample mean of the outcome. 
For a normally distributed outcome, this is exact as $v(\cdot)=1$. 
For a binary outcome, since population stratification is the primary confounder of the 
SNP-outcome relationship in GWAS, and because $\mu_{i0}$ generally varies slowly with the 
principal components, this approximation is practically reasonable. 
Ultimately we are approximating $\mathbf{G}_{.j}^{T}\mathbf{P}\mathbf{G}_{.k}$ in 
(\ref{p2_eq:estimate_cov}) by $\mathbf{G}_{.j}^{(r)T}\mathbf{G}_{.k}^{(r)}-\mathbf{G}_{.j}^{(r)T}\mathbf{X}^{(r)}
\{\mathbf{X}^{(r)T}\mathbf{X}^{(r)}\}^{-1}\mathbf{X}^{(r)T}\mathbf{G}_{.k}^{(r)}$ up to a scale 
parameter, with the scale parameter eventually cancelled out in $\widehat{\Sigma}_{jk}$.

%%%%%%%%%%%%%%%%%%%%%%%%%%%%%%%%%%%%%%%%%%%%%%%%%%%%%%%%%%%%
%%%%%%%%%%%%%%%%%%%%%%%%%%%%%%%%%%%%%%%%%%%%%%%%%%%%%%%%%%%%
%%%%%%%%%%%%%%%%%%%%%%%%%%%%%%%%%%%%%%%%%%%%%%%%%%%%%%%%%%%%
\section{The Generalized Berk-Jones Test for SNP-Set Effects}
\label{p2_sec:GBJ}

\subsection{The Berk-Jones Statistic}
\label{p2_ss::BJ}
We briefly review the Berk-Jones statistic in this section to help introduce the Generalized 
Berk-Jones statistic in Section \ref{p2_ss:GBJ}. 
The BJ statistic is designed to test for $H_0: \boldsymbol{\beta}=0$  against the alternative 
that a nonempty subset of the $\beta_j$ are nonzero, assuming the marginal test statistics 
are independent.
Let $\bar{\Phi}(t)=1-\Phi(t)$ denote the survival function of a standard normal random variable 
and $\Phi^{-1}(t)$ denote its inverse.
Let $|Z|_{(j)}$ denote the order statistics of the vector that results from applying the absolute 
value operator to each element of $\mathbf{Z}$, so that $|Z|_{(1)}$ is the smallest value of 
$\mathbf{Z}$ in magnitude. 

Set $S(t)=\sum_{j=1}^{d}\mathbf{1}\left(|Z_{j}|\geq t\right)$, which is the number of marginal 
test statistics with a magnitude greater than or equal to some threshold $t$.  
For a fixed $t\geq0$, and if $Z_{j}\overset{\text{iid}}{\sim}N(0,1)$ for all $j$, then $S(t)$ has
a binomial distribution with size $d$ and mean parameter $\pi=2\bar{\Phi}(t)$. 
This viewpoint motivates the Berk-Jones statistic for independent observations \citep{HC} as:
\begingroup
	\makeatletter\def\f@size{11}\check@mathfonts
	\begin{eqnarray}
	BJ_{d}
 	& = & \max_{t \geq |Z|_{(d/2+1)}}\left[S(t)\log\left\{ \frac{S(t)}{2d\bar{\Phi}(t)}\right\} +\left\{ d-S(t)\right\} \log\left\{ \frac{1-S(t)/d}	{1-2\bar{\Phi}(t)}\right\} \right]\mathbf{1}\left\{ 2\bar{\Phi}(t)<\frac{S(t)}{d}\right\} \label{p2_eq:bj}\\
  	& = & \max_{1\leq j\leq d/2}\log\left[\frac{\text{Pr}\left\{ S(|Z|_{(d-j+1)})=j\bigg| E(\mathbf{Z}) = \widehat{\mu}_{j,d} \cdot \mathbf{J}_{d} \right\} }{\text{Pr}\left\{ S(|Z|_{(d-j+1)})=j\bigg| E(\mathbf{Z}) = 0 \cdot \mathbf{J}_{d} \right\} }\right]\mathbf{1}\left\{2\bar{\Phi}\left(|Z|_{(d-j+1)}\right)<\frac{j}{d}\right\}  \nonumber 
	\end{eqnarray}
\endgroup
where $\mathbf{J}_{d}^{T} = (1,1,...,1)_{1\times d}$,  $\widehat{\mu}_{j,d}>0$ solves the equation 
\begin{equation}
j/d = 1 - \left\{ \Phi(|Z|_{(d-j+1)}-\widehat{\mu}_{j,d})-\Phi(-|Z|_{(d-j+1)}-\widehat{\mu}_{j,d})\right\},
\label{p2_eq:mu_Z}
\end{equation}
and the second line uses the characterization of $S(t)$ as a binomial random variable.

We see that BJ can roughly be explained as the maximum of a one-sided likelihood ratio test
on the mean parameter of $S(t)$, performed over the larger half of observed test statistic magnitudes.
At $t=|Z|_{(d-j+1)}$ we have $\pi=2\bar{\Phi} \left(|Z|_{(d-j+1)}\right)$ under the binomial likelihood null, and 
we have $\pi=j/d$ under the binomial likelihood alternative.
We say binomial likelihood null and alternative to make clear that we are talking about an 
interpretation of the Berk-Jones statistic and to distinguish from the actual set-based null and 
alternative hypotheses being tested.

Note that the Higher Criticism test differs from the Berk-Jones by replacing the likelihood 
ratio statistic in (\ref{p2_eq:bj}) with the Pearson Chi-square statistic $\{S(t)-2d\bar\Phi(t)\}^2/\{2d\bar\Phi(t)(1-2\bar\Phi(t)\}$. 
Let $k$ be the number of causal SNPs in a set. 
The sparse regime is designated as $k<d^{1/2}$, and we call $d^{1/4}<k<d^{1/2}$ moderately 
sparse, with $k\leq d^{1/4}$ referred to as extremely sparse.
\citet{HC} showed that, when the $Z_{j}$ are all mutually independent, both HC and BJ are able to 
reach the detection boundary over the entire sparse signal regime as $d\rightarrow\infty$.  
\citet{ALR} and \citet{LiSiegmund}  showed that the BJ statistic generally has better 
power than HC when the size of the set  $d$ is finite and the signals are moderately sparse.

If $Z_{1},...,Z_{d}$ are correlated, as they will be for test statistics arising from neighboring 
SNPs in a gene, then $S(t)$ no longer has a binomial distribution under the null. 
In this case, the standard Berk-Jones statistic no longer has a meaningful interpretation, 
and we may expect it to lose efficiency. 
In fact, we will show later that the rejection region of the Berk-Jones has a less desirable 
shape under various correlation structures, leading to a significant loss in power when 
the test statistics are not independent. 
Therefore we are interested in developing a modified BJ statistic that can account for 
correlation among the marginal test statistics in a set and thus possesses rejection regions 
which are more robust to arbitrary correlation structures. 

%%%%%%%%%%%%%%%%%%%%%%%%%%%%%%%%%%%%%%%%%%%%%%%%%%%%%%%%%%%%
%%%%%%%%%%%%%%%%%%%%%%%%%%%%%%%%%%%%%%%%%%%%%%%%%%%%%%%%%%%%
%%%%%%%%%%%%%%%%%%%%%%%%%%%%%%%%%%%%%%%%%%%%%%%%%%%%%%%%%%%%
\subsection{The Generalized Berk-Jones Statistic}
\label{p2_ss:GBJ}

%\subsection{Definition of the Generalized Berk-Jones Test Statistic}

We now propose the Generalized Berk-Jones statistic for testing the association between 
a SNP-set and outcome.   
Following the spirit of Berk-Jones, GBJ considers a likelihood ratio type statistic on 
the  mean parameter of $S(t)$, but the key difference is GBJ explicitly accounts for the 
correlation  structure of $Z_{1},...,Z_{d}$. 
More precisely, we define the GBJ statistic as:
\begin{eqnarray*}
GBJ_{d}&=&\max_{1\leq j\leq d/2 }\log\left[\frac{\text{Pr}\left\{ S\left(|Z|_{(d-j+1)}\right)=j\bigg| E(\mathbf{Z})=\widehat{\mu}_{j,d} \cdot \mathbf{J}_{d},\text{cov}(\mathbf{Z})=\boldsymbol{\Sigma}\right\} }{\text{Pr}\left\{ S\left(|Z|_{(d-j+1)}\right)=j\bigg| E(\mathbf{Z})=0 \cdot \mathbf{J}_{d},\text{cov}(\mathbf{Z})=\boldsymbol{\Sigma}\right\} }\right] \\
 & & \cdot \mathbf{1}\left\{2\bar{\Phi}\left(|Z|_{(d-j+1)}\right)<\frac{j}{d}\right\}. 
\end{eqnarray*} 

When the $Z_j$ are correlated, $S(t)$ follows either an underdispersed or overdispersed binomial distribution 
instead of the standard binomial. 
However finding the exact distribution of $S(t)$ when $\boldsymbol{\Sigma}\neq\mathbf{I}$ 
is difficult.
For a general $\boldsymbol{\Sigma}$, computing $\text{Pr}\left\{S(t)=m\right\}$ requires 
iterating through $d$ choose $m$ terms and is very time consuming. 
In special cases, such as when $\boldsymbol{\Sigma}$ has an exchangeable correlation 
structure with $\boldsymbol{\Sigma}=(1-\rho)\mathbf{I} + \rho\mathbf{1}\mathbf{1}^{T}$, 
the calculation is much easier. 
However these scenarios occur rarely, if ever, in practice.

We propose to approximate the full distribution of $S(t)$ using an Extended Beta-Binomial 
(EBB) distribution \citep{EBB}. 
The Extended Beta-Binomial is a reparameterization and extension of the standard 
Beta-Binomial$(\alpha,\beta)$ distribution with the standard Beta-Binomial being a special 
case of the EBB. 
A random variable $V\sim\text{EBB}\left(d,\lambda,\gamma\right)$ has probability mass function
\begin{equation}
\text{Pr}\left(V=v;d,\lambda,\gamma\right)=\left(\begin{array}{c}
d\\
v
\end{array}\right) \prod_{k=0}^{v-1}(\lambda+\gamma k)\prod_{k=0}^{d-v-1}(1-\lambda+\gamma k) \bigg/ \prod_{k=0}^{d-1}(1+\gamma k),
\label{p2_eq:EBB_PMF}
\end{equation}
where we follow the convention $\prod_{k=0}^{a}c_{k}=1$ for $a<0$.
The mean of $V$ is given by $E(V)=d\lambda$ and the variance is 
$\text{Var}(V)=d\lambda(1-\lambda)\left\{1+(d-1)\gamma(1+\gamma)^{-1}\right\}$.

The Extended Beta-Binomial distribution reduces to the Beta-Binomial distribution if we set
$\lambda=\alpha(\alpha+\beta)^{-1}$ and 
$\gamma=(\alpha+\beta)^{-1}$ for $\alpha,\beta>0$.
Because the standard Beta-Binomial distribution requires $\alpha,\beta>0$, it cannot 
accommodate underdispersion and never reduces to the binomial distribution. 
In contrast, the EBB allows for both overdispersion and underdispersion, and it reduces 
exactly to the binomial distribution when $\gamma=0$. 
This mechanism allows our GBJ statistic to reduce to the Berk-Jones when there is no 
correlation among the observations.

\subsection{Calculation of the Generalized Berk-Jones Statistic}
\label{p2_ss:calculate_GBJ}

We now describe more precisely the mechanics of calculating the GBJ
statistic. To begin, check if the condition 
$\mathbf{1}\left\{2\bar{\Phi}\left(|Z|_{(d-j+1)}\right) < j/d \right\}$ is satisfied at 
any $j \leq d/2$.
If the condition is never satisfied, then the observed value of GBJ is 0
and we do not need to perform any more computation. 
The following steps should only be taken on indices $j \leq d/2 $ where the 
condition is satisfied.

At each qualifying $j$, we  approximate the distribution of $S(|Z|_{(d-j+1)})$ 
by an Extended Beta-Binomial random variable under both the binomial 
likelihood null and the binomial likelihood alternative.
Denote these two variables by $V_{0}^{(j)}\sim EBB(\lambda_{0}^{(j)},\gamma_{0}^{(j)})$
and $V_{a}^{(j)}\sim EBB(\lambda_{a}^{(j)},\gamma_{a}^{(j)})$. 
We solve for $(\lambda_{0}^{(j)},\gamma_{0}^{(j)})$ through moment matching equations 
\begin{eqnarray*}
\lambda_{0}^{(j)} & = & E_{0}\left\{ S(|Z|_{(d-j+1)})\right\} /d, \\
\frac{\gamma_{0}^{(j)}}{1+\gamma_{0}^{(j)}} & = & \frac{\text{Var}_{0}\left\{ S(|Z|_{(d-j+1)})\right\} -d\lambda_{0}^{(j)}(1-\lambda_{0}^{(j)})}{d(d-1)\lambda_{0}^{(j)}(1-\lambda_{0}^{(j)})},
\end{eqnarray*}
where $E_{0}$ and $\text{Var}_{0}$ denote the expectation and variance
conditional on $\mathbf{Z}\sim MVN(0 \cdot \mathbf{J}_{d},\boldsymbol{\Sigma})$.
Similarly, we solve for $(\lambda_{a}^{(j)},\gamma_{a}^{(j)})$ using
the same equations except with $E_{a}$ and $\text{Var}_{a}$, which
are the expectation and variance conditional on $\mathbf{Z}\sim MVN(\widehat{\mu}_{j,d} \cdot \mathbf{J}_{d},\boldsymbol{\Sigma})$.

\sloppy
The first moment matching equation is simple to solve, since clearly
$E_{0}\left\{ S(|Z|_{(d-j+1)})\right\} = 2d\bar{\Phi}(|Z|_{(d-j+1)})$
and $E_{a}\left\{ S(|Z|_{(d-j+1)})\right\} = j$. 
The variance term in the second equation is more difficult. 
We can use Theorem 1 of \citet{GHC} for $\text{Var}_{0}$.
For $\text{Var}_{a}$, we need the following theorem:

\fussy
\begin{theorem} Let  $\mathbf{Z}\sim MVN( \mu \cdot \mathbf{J}_{d},\boldsymbol{\Sigma})$,
and take $\mathcal{H}_{i}(t)$ be the probabilists' Hermite polynomials.
Define $\bar{\rho^{r}}=\frac{2}{d(d-1)}\sum_{1\leq k<l\leq d}\left(\rho_{k,l}\right)^{r}$, where
$\rho_{k,l}$ is the $(k,l)$ element of $\boldsymbol{\Sigma}$, and let $\rho_{k,k}=1$ for all $k=1,2,..,d$. Then the variance of $S(t)$ is:
\begin{eqnarray*}
\text{Var}_{a}\left\{S(t)\right\} & = & d \lambda (1-\lambda) + d(d-1)\left\{ \phi(t-\mu)^{2}\sum_{r=1}^{\infty}\frac{\bar{\rho^{r}}}{r!}H_{r-1}(t-\mu)^{2}\right\} \\
 &  & +d(d-1)\left\{ \phi(-t-\mu)^{2}\sum_{r=1}^{\infty}\frac{\bar{\rho^{r}}}{r!}H_{r-1}(-t-\mu)^{2}\right\} \\
 &  & -2d(d-1)\left\{ \phi(-t-\mu)\phi(t-\mu)\sum_{r=1}^{\infty}\frac{\bar{\rho^{r}}}{r!}H_{r-1}(-t-\mu)H_{r-1}(t-\mu)\right\}, \\
 \lambda & = & 1 - \left \{ \Phi(t-\mu) - \Phi(-t-\mu) \right \}.
\end{eqnarray*}
\end{theorem}
The proof of this theorem is given in the Supplementary Materials. 
The terms in the infinite sum shrink very quickly, and in practice we see good accuracy
using only the first ten.

After matching all four parameters $(\lambda_{0}^{(j)},\gamma_{0}^{(j)}, \lambda_{a}^{(j)},\gamma_{a}^{(j)})$,
we calculate 
\[
GBJ_{d}^{(j)}=\log\left\{ \frac{\text{Pr}\left(V_{a}^{(j)}=j;d,\lambda_{a}^{(j)},\gamma_{a}^{(j)}\right)}{\text{Pr}\left(V_{0}^{(j)}=j;d,\lambda_{0}^{(j)},\gamma_{0}^{(j)}\right)}\right\}. 
\]
The maximum value of $GBJ_{d}^{(j)}$ among all qualifying $j$ is then the 
observed Generalized Berk-Jones statistic. 

%%%%%%%%%%%%%%%%%%%%%%%%%%%%%%%%%%%%%%%%%%%%%%%%%%%%%%%%%%%%
%%%%%%%%%%%%%%%%%%%%%%%%%%%%%%%%%%%%%%%%%%%%%%%%%%%%%%%%%%%%
%%%%%%%%%%%%%%%%%%%%%%%%%%%%%%%%%%%%%%%%%%%%%%%%%%%%%%%%%%%%
\subsection{Analytic P-value Calculation}
\label{p2_ss:pvalue}

Let $G_{d}$ be a general supremum-based global statistic such as the GBJ statistic.
Suppose $G_d$ is constructed using independent marginal test statistics $Z_{1},...,Z_{d}$.
Denote the observed value of this statistic by $g$, where higher
values of $g$ indicate more evidence for the alternative.  
As noted by  \citet{boundary_crossing}, the p-value for $g$ can often be written
\[
\text{Pr}\left(G_{d}\geq g\right)=1-\text{Pr}\left\{ \forall j=1,2,...,d:|Z|_{(j)}\leq b_{j}\bigg|Z_{j}\overset{iid}{\sim}N(0,1)\right\}, 
\]
where $0\leq b_{1}\leq b_{2}\leq...\leq b_{d}$ are 'boundary points' that come from 
inversion of the test statistic.  
The points $b_{1},b_{2},...,b_{d}$ will depend on $g$ and $d$ and will be different for
different choices of a global statistic, but for the sake of presentation, 
we will suppress this dependency in the notation.
\citet{boundary_crossing} proposed a method that can calculate the p-value of $G_d$ very 
quickly if $Z_{1},...,Z_{d}$ are independent. 
However when $Z_{1},...,Z_{d}$ are correlated, their techniques for a fast calculation are not applicable. 

An exact p-value for GBJ, and for any global test applied to correlated
observations, must take into account the covariance structure of $\mathbf{Z}$.
The p-value for these tests is then
\begin{eqnarray}
\text{Pr}\left(G_{d}\geq g\right) & = & 1-\text{Pr}\left\{ \forall j=1,2,...,d:|Z|_{(j)}\leq b_{j}\bigg|\mathbf{Z}\sim MVN(\mathbf{0},\boldsymbol{\Sigma})\right\},
\label{p2_eq:exact_pvalue}
\end{eqnarray}
where $b_{j}$ is understood to additionally depend on $\boldsymbol{\Sigma}$ as well.
We are unaware of any computationally feasible expressions to calculate the joint distribution
of the order statistics $|Z|_{(1)},...,|Z|_{(d)}$ when $d$ is moderate or large. 
The Supplementary Materials provides a procedure to compute this probability by 
partitioning the region into $d!$ separate sections, but the method is very 
computationally intensive and not feasible for use with $d>10$.

However, an alternative way to write the rejection region of equation (\ref{p2_eq:exact_pvalue}) is
\begingroup
	\makeatletter\def\f@size{11}\check@mathfonts
	\begin{equation}
	\text{Pr} \left\{ \forall j:|Z|_{(j)}\leq b_{j}\bigg|\mathbf{Z}\sim MVN(\mathbf{0},\boldsymbol{\Sigma})\right\} =\text{Pr}\left\{ \forall j:S(b_{j})\leq(d-j)\bigg|\mathbf{Z}\sim MVN(\mathbf{0},\boldsymbol{\Sigma})\right\}.
	\label{p2_eq:alt_exact_pvalue}
	\end{equation}
\endgroup
The right hand side of (\ref{p2_eq:alt_exact_pvalue}) suggests that the quantity can be calculated recursively.  
Indeed, define $b_{0}=0$ and $q_{j,a}=\text{Pr}\left\{ S(b_{j})=a,\bigcap_{k=1}^{j-1}S(b_{k})\leq d-k\right\}$.  
The quantity in (\ref{p2_eq:alt_exact_pvalue}) is just $q_{d,0}$ and can be calculated recursively as
\begin{eqnarray}
q_{j,a}  & = & \sum_{m=a}^{d-j+1}\text{Pr}\left\{ S(b_{j})=a,S(b_{j-1})=m,\bigcap_{k=1}^{j-2}S(b_{k})\leq d-k\right\}, \nonumber\\
 & = & \sum_{m=a}^{d-j+1}\text{Pr}\left\{ S(b_{j})=a \bigg| S(b_{j-1})=m,\bigcap_{k=1}^{j-2}S(b_{k})\leq d-k\right\} q_{j-1,m}, \nonumber\\
 & \approx & \sum_{m=a}^{d-j+1}\text{Pr}\left\{ S(b_{j})=a \bigg| S(b_{j-1})=m\right\} q_{j-1,m},
 \label{p2_eq:q_ja}
\end{eqnarray}
for $j>1$, with $q_{1,a}=\text{Pr} \left\{S(b_{1})=a | S(b_{0})=d\right\}$. We use an EBB approximation for the distribution of $S(b_{j})$ conditional on $S(b_{j-1})=m$, with the equations
\begin{eqnarray*}
\lambda_{j} & = & \frac{\bar{\Phi}(b_{j})}{\bar{\Phi}(b_{j-1})}, \\
\frac{\gamma_{j}}{1+\gamma_{j}} & = & \frac{2\sum_{k<l}\left[\frac{\text{Pr}\left( |Z_{k}|,|Z_{l}|\geq b_{j}\right) }{\text{Pr}\left( |Z_{k}|,|Z_{l}|\geq b_{j-1}\right) }-\left\{ \frac{\bar{\Phi}(b_{j})}{\bar{\Phi}(b_{j-1})}\right\} ^{2}\right]}{d(d-1)\lambda_{j}(1-\lambda_{j})},
\end{eqnarray*}
to match the moments.  
Finally, set $\text{Pr}\left\{ S(b_{j})=a|S(b_{j-1})=m\right\} := 
\text{Pr}\left(V_{j}=a\right)$ where $V_{j}\sim EBB(m, \lambda_{j},\gamma_{j})$.
Evaluation of $\text{Pr}\left(|Z_{k}|,|Z_{l}|\geq b_{j}\right)$ follows from steps similar to the proof of Theorem 1.

Note that we can generalize the scheme described above to calculate
p-values for many different supremum-based global tests by adopting the
general approach of \citet{boundary_crossing}. 
As long as the test statistic can be inverted to create the bounds $b_{1},...,b_{d}$, 
we can use equation (\ref{p2_eq:q_ja}) to calculate its p-value when applied to correlated observations.  
In particular, we can use this procedure to perform p-value calculations for the  
HC, GHC, BJ, and GBJ statistics.
Both the calculation and the Generalized Berk-Jones test are implemented in the 
R package \verb|GBJ|, available on the \verb|CRAN| repository.

%%%%%%%%%%%%%%%%%%%%%%%%%%%%%%%%%%%%%%%%%%%%%%%%%%%%%%%%%%%%
%%%%%%%%%%%%%%%%%%%%%%%%%%%%%%%%%%%%%%%%%%%%%%%%%%%%%%%%%%%%
%%%%%%%%%%%%%%%%%%%%%%%%%%%%%%%%%%%%%%%%%%%%%%%%%%%%%%%%%%%%
\subsection{The Omnibus Test}
\label{p2_ss:omnibus}

While we will show that the Generalized Berk-Jones test possesses an attractive finite sample
rejection region when signals are moderately sparse, GBJ may also lose power in the presence 
of very sparse or dense signals.  
As SNP-set inference involves testing for a composite alternative $H_1:\boldsymbol{\beta} \neq 0$, 
there is no uniformly optimal test for both sparse and dense alternatives. 
Since signal sparsity varies between genes, the best test will also change from gene to gene, but it is unknown prior to 
scanning the genome.  
Thus we propose an omnibus test that offers robust power over a range of 
different sparsity levels.

The omnibus test is constructed by combining the SKAT, GBJ, GHC, 
and minimum p-value statistics, which have been described above. 
The motivation for choosing these four methods is to combine tests that are known 
to have good power  when signals are dense, moderately sparse, very sparse, 
and the sparsest possible, respectively.  
The MinP method uses the set's largest marginal test statistic in magnitude $|Z|_{(d)}$ 
as a test statistic. 
When the $Z_{j}$ are independent, \citet{HC} showed that MinP asymptotically 
reaches the same detection boundary as HC and BJ in the very sparse regime 
$k \leq d^{1/4}$ but not the moderately sparse regime $d^{1/4}<k<d^{1/2}$.
In finite samples, MinP can have better power than the other three tests when there 
are only one or two causal SNPs.
In contrast, SKAT is known to have high power when signals in a SNP-set are dense.

The omnibus test first applies each of the four tests to the same SNP-set, and then it 
carries forward the smallest p-value from the four tests as a test statistic. 
Specifically, the omnibus test statistic is defined as: 
\[
OMNI=\min\left(p_{GBJ},p_{GHC},p_{SKAT},p_{MinP}\right ),
\]
where $p_{GBJ}, p_{GHC}, p_{SKAT}$, and $p_{MinP}$  denote the p-values of the four 
tests applied on the same SNP-set.
As these tests are applied to the same data, the four p-values will be correlated.

Calculations of the p-value for OMNI must again account for the correlation between tests. 
We employ a Gaussian copula approximation for the joint distribution of the inverse-normal transformed p-values:
\[
p_{OMNI}=1-\Phi_M\left[ \left \{ \Phi^{-1}(1-OMNI),...,\Phi^{-1}(1-OMNI) \right \}_{4\times 1};\mathbf{R}_{4\times4} \right],
\]
where $\Phi_M(\cdot;\mathbf{R})$ denotes the  joint cumulative distribution function 
of a multivariate  normal distribution  with mean vector zero and correlation matrix $\mathbf{R}$.   
The correlation matrix $\mathbf{R}$ of the four component test statistics is estimated through 
parametric bootstrap under the null. 
For each subject $i$ in the study, we simulate a new outcome based on the null 
mean $\widehat{\mu}_{0i}$.  
When individual-level data are not available, we take $\widehat{\mu}_{0i}$ to be the same 
constant for all subjects as an approximation.
Then each of the four tests are applied with the simulated outcome instead of the original one.
The original design matrix, or approximated design matrix if working with summary statistics, 
is used each time.
Each of the four p-values is subtracted from 1 and then inverse-normal transformed; under the null hypothesis the 
four transformed values have marginal normal distributions with mean zero.
As we only need to estimate the correlation matrix $\mathbf{R}$, only a small number
of parametric bootstrap samples are needed.
In practice, this procedure is repeated 100 times, and then we set $\mathbf{R}$  equal to the sample 
correlations of the inverse-normal transformed statistics.  
We will see that this omnibus test performs well across a variety of settings.

%%%%%%%%%%%%%%%%%%%%%%%%%%%%%%%%%%%%%%%%%%%%%%%%%%%%%%%%%%%%
%%%%%%%%%%%%%%%%%%%%%%%%%%%%%%%%%%%%%%%%%%%%%%%%%%%%%%%%%%%%
%%%%%%%%%%%%%%%%%%%%%%%%%%%%%%%%%%%%%%%%%%%%%%%%%%%%%%%%%%%%
\section{Rejection Region Analysis of Different SNP-Set Tests}
\label{p2_sec:rej_region}

We study in this section the finite sample rejection regions for the BJ, GBJ, HC, and GHC tests,
and we advocate for viewing these statistics as boundary-defining algorithms. 
Consider a fixed SNP-set $s$ with size $d=d_{s}$ and correlation structure $\mathbf{\Sigma}=\mathbf{\Sigma}_{s}$, and 
suppose we wish to conduct inference at level $\alpha=0.01$. 
Using the p-value calculation from above, we can employ standard 
root-finding routines to find the observed value $g_{s}$ which 
would result in a GBJ p-value of 0.01 for a SNP-set with these characteristics.
Then inverting $g_{s}$ to find boundary points $b_{1},b_{2},...,b_{d_{s}}$
as in Section~\ref{p2_ss:pvalue} constructs a rejection region in terms of in terms of $|Z|_{(1)},...,|Z|_{(d_{s})}$.
That is, if the observed value of $|Z|_{(j)}$ were larger than $b_{j}$ for any $j=1,2,...,d_{s}$, then the 
GBJ p-value for this set would be less than $0.01$.
Plotting the bounds for various tests using different values of $\alpha$, $d$, and $\boldsymbol{\Sigma}$
shows us exactly what types of signals a given test is well-powered to detect at 
level $\alpha$.
For the same setting, a test with smaller bounds is preferred, as it will provide more
finite sample power.

To numerically compare the rejection boundaries, consider a simplified model of SNP-set correlation
structure where the set is partitioned into only two sections. 
Let one section be the independence section, where all SNPs in this portion
are completely independent of all other SNPs in the set.  
Let the other section be the correlated section, where all SNPs in this portion
have common pairwise correlation $\rho$ with other SNPs in the section.
For our numerical study, $\rho=0.3$ for the correlated section.
We investigate SNP-sets of size $d=20$ and $100$,
correlated sections which contain 50\% and 75\% of the SNPs,
and tests at size $\alpha=0.01$. 
These parameters are chosen to represent reasonable boundaries on the correlation structures seen 
in common GWAS data; \citet{Dawson_LD} estimated that the average $r^{2}$ between 
SNPs separated by 100kb is around 0.1.

\begin{figure}[!ht]
\begin{center}
\centerline{\includegraphics[scale=0.25, angle=270]{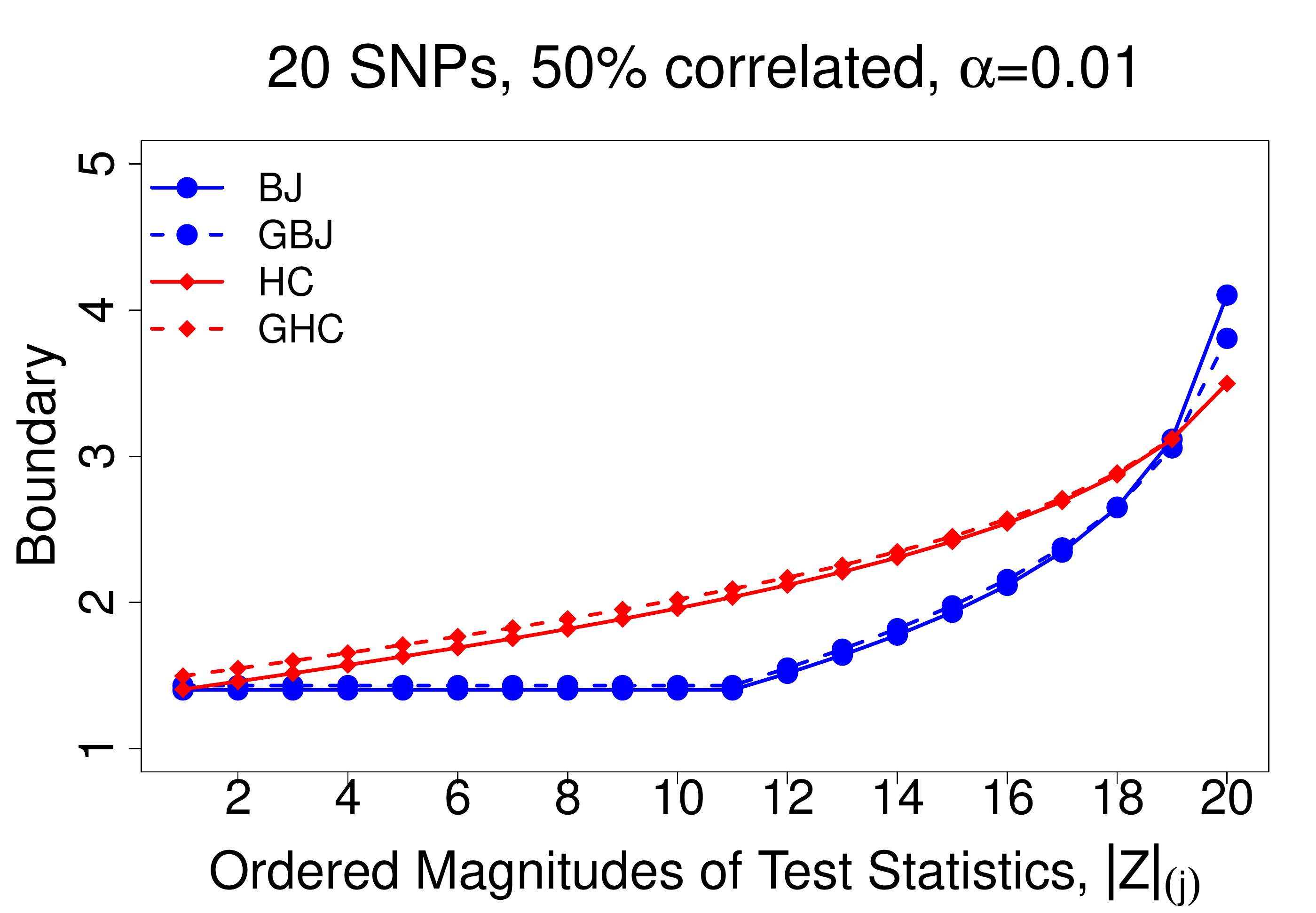}\includegraphics[scale=0.25, angle=270]{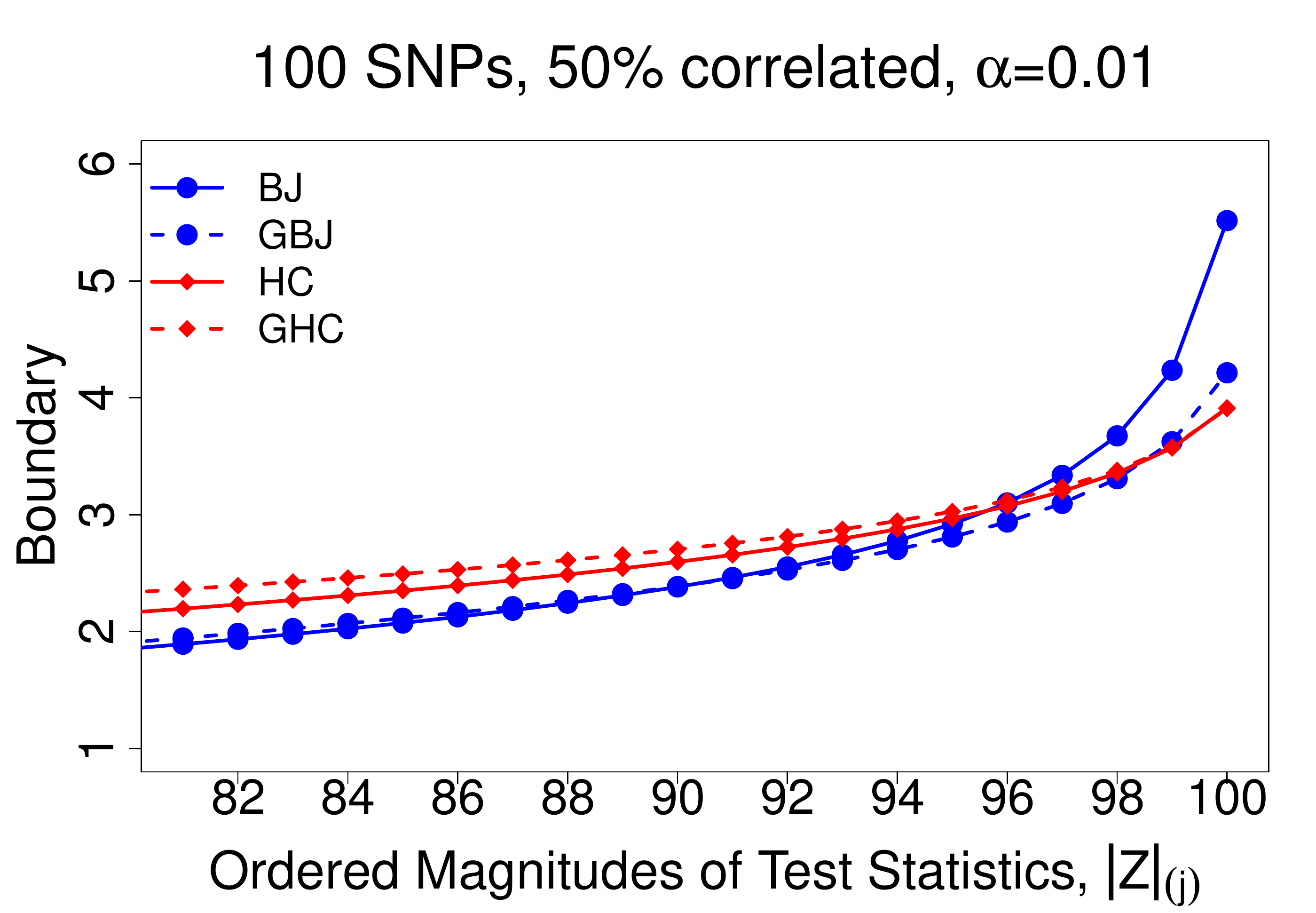}}
\centerline{\includegraphics[scale=0.25, angle=270]{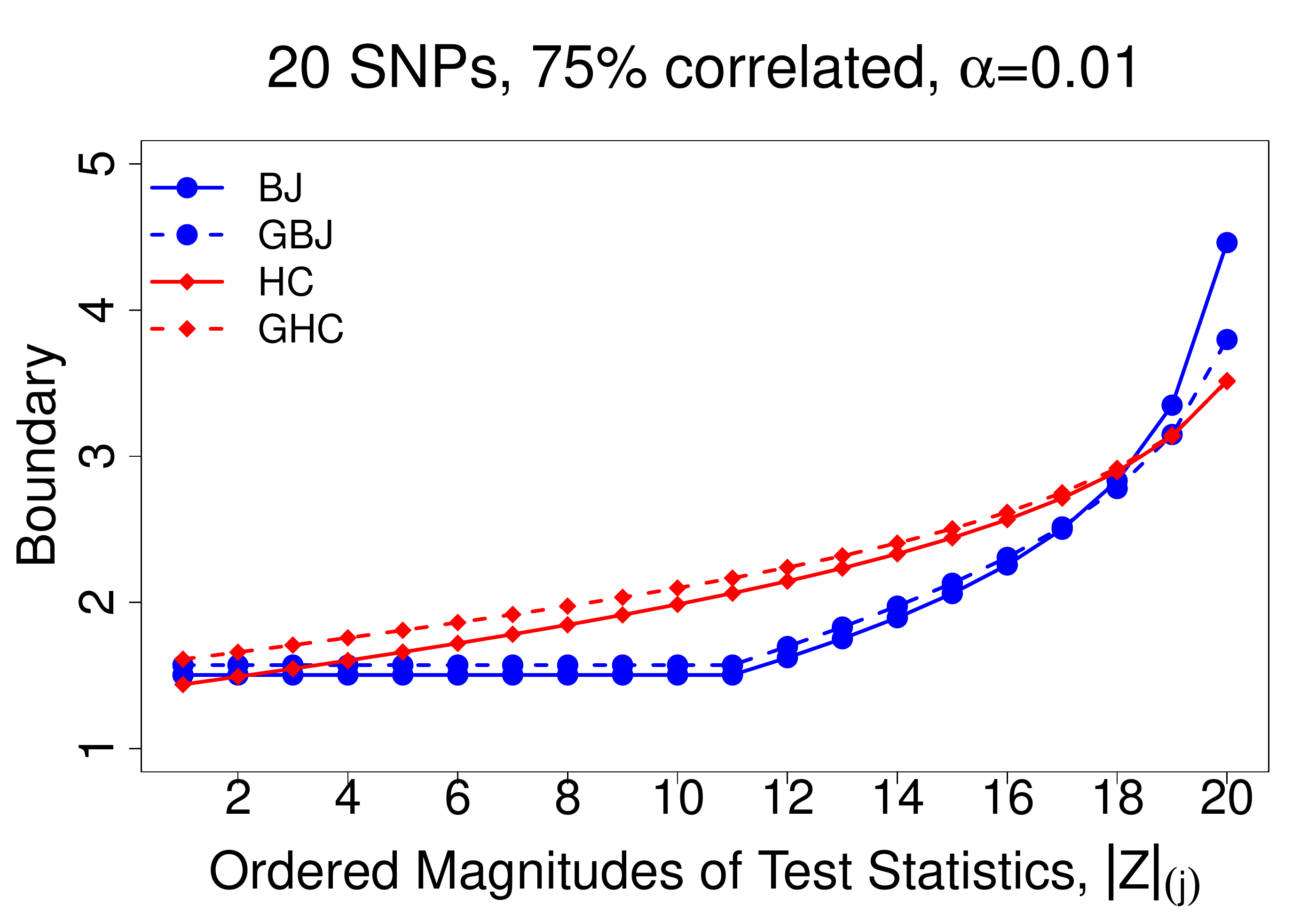}\includegraphics[scale=0.25, angle=270]{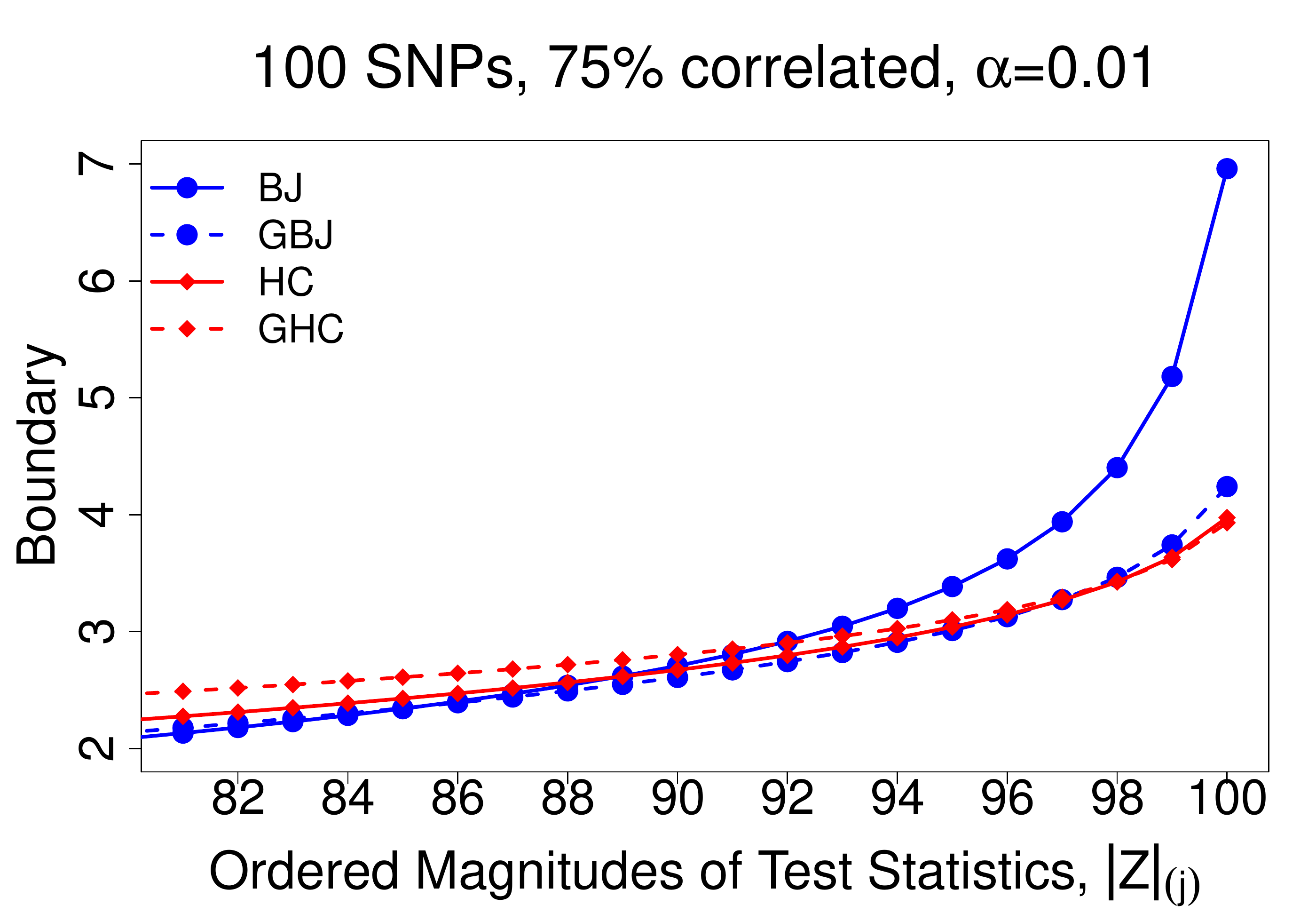}}
\end{center}
\caption{Rejection region of Berk-Jones, Generalized Berk-Jones, Higher Criticism, and Generalized Higher Criticism tests, plotted according to the order statistics of the absolute values of the test statistics.  If the $j$th smallest test statistic in magnitude is greater than the boundary point for a given test at any point $j$ on the x-axis, then we would reject the null using that test at level $\alpha=0.01$. The difference between BJ and GBJ becomes much more pronounced as both the size of the set and the amount of correlation increase.}
\label{p2_fig:rej_region_fig}
\end{figure}

The rejection regions for each SNP-set are plotted in Figure \ref{p2_fig:rej_region_fig}. 
At the $j$th coordinate on the x-axis, if the observed $|Z|_{(j)}$ lies above the boundary
of a particular test at that coordinate, then we would reject the
null for that test at level $\alpha=0.01$. 
The lines on the graph are added to aid in visualization, but there should 
be no interpretation of interpolation between the points. 
It does not make sense to think of the boundary at $|Z|_{(2.5)}$, for example.
While standard methods for inference on HC and BJ are invalid in the presence of
correlation, valid p-values for these tests can be computed with the same ideas
we have introduced for GBJ inference, specifically following equations (\ref{p2_eq:exact_pvalue})-(\ref{p2_eq:q_ja}).
Thus we can show that HC and BJ sometimes have much less desirable 
rejection regions when SNPs in a set are correlated.

%%%%%%%%%%%%%%%%%%%%%%%%%%%%%%%%%%%%%%%%
%%%%%%%%%%%%%%%%%%%%%%%%%%%%%%%%%%%%%%%%
%%%%%%%%%%%%%%%%%%%%%%%%%%%%%%%%%%%%%%%%
%%%%%%%%%%%%%%%%%%%%%%%%%%%%%%%%%%%%%%%%

One of the clearest trends from Figure \ref{p2_fig:rej_region_fig} is that the HC and GHC boundaries
are lower for a small region around $|Z|_{(d)}$, and then the BJ
and GBJ boundaries quickly become smaller as we move left. 
This behavior indicates that HC and GHC are better at detecting the sparsest
alternatives with only a very few signals, as those signals would
almost always manifest as the test statistics with the largest magnitude.
In contrast, the plots demonstrate that BJ and GBJ can have more power
to detect weaker, less sparse signals which may be more easily found
by examining the test statistic which is, say, fifth or tenth largest
in magnitude. 
The boundaries of HC and GHC can drop below BJ and GBJ again for the smallest
observed magnitudes, but signals would only be found in these observations if they 
are particularly dense, a setting which is not the focus of our efforts. 
The intuition we can glean from this figure is closely aligned with the
theoretical development of \citet{HC} and the simulations
of \citet{LiSiegmund} when the marginal test statistics $Z_j$ are independent. 
These authors showed that HC is attuned to
detect sparse signals arising at the very tail of the observed
distribution, while BJ has more power as the number of signals rises.

These results also show why BJ is likely to have low power for detecting sparse signals
when the level of correlation is high. 
When 75\% of the SNPs are correlated, the rejection boundary for BJ at the largest few observations
is the highest by multiple orders of magnitude on the p-value scale.
It would not be desirable to apply BJ in these types of settings,
as the test loses an extremely large amount of sensitivity to detect
signals in the most outlying values. 
BJ is still likely to be suitable for detecting dense signals in these situations.
Here, GBJ acts as a compromise between BJ and GHC under high correlation.
GBJ provides a much lower boundary than BJ at the tail in exchange for slightly
higher boundaries near the middle. 
Thus, GBJ can detect both sparse and dense signals in this example.
On the other hand, GBJ provides a slightly higher boundary than GHC at the tail in 
exchange for lower boundaries past the tail, so it trades some power in the 
extremely sparse regime for more power to detect moderately sparse signals.

We see that choosing a different statistic is essentially choosing a different boundary-setting 
algorithm, and this choice should ideally be informed by parameters such as the amount of 
correlation and estimated sparsity level.
Ultimately these plots illustrate that there is no single best global test for all types
of alternatives. 
A genome-wide analysis strategy using the omnibus test will be likely to 
have robust power across different sparsity settings, correlation
structures, and SNP-set sizes.

%%%%%%%%%%%%%%%%%%%%%%%%%%%%%%%%%%%%%%%%%%%%%%%%%%%%%%%%%%%%
%%%%%%%%%%%%%%%%%%%%%%%%%%%%%%%%%%%%%%%%%%%%%%%%%%%%%%%%%%%%
%%%%%%%%%%%%%%%%%%%%%%%%%%%%%%%%%%%%%%%%%%%%%%%%%%%%%%%%%%%%
\section{Simulation Results}
\label{p2_sec:simulation}

\subsection{Type I Error of the Generalized Berk-Jones Test}
\label{p2_sss:TypeIerr}

We first illustrate that our p-value calculations for the GBJ and omnibus tests are accurate enough to control
the Type I error rate at levels required to declare genome-wide significance of a SNP-set. 
To replicate the setting of traditional GWAS data, we perform the size simulation 
on random regions across chromosome 5 which correspond to known gene sizes.
We also conduct the size simulation on a high-LD subset of the FGFR2 gene
and a low-LD subset of the FGFR2 gene to parse LD-related effects.
We choose FGFR2 because it contains both high and low LD regions,
and because it will later be the most significant gene in our analysis
of the CGEMS data.
All SNP-sets contain genotypes simulated with HAPGEN2 \citep*{HAPGEN2} using the CEU population 
from HapMap3 as a reference panel.

In all simulations the outcome is generated as $Y\sim N(0,1)$, and
we fit the linear regression model (1) with $\boldsymbol{\beta}=0$
and $\mathbf{X}_{i}=1$. 
Each simulation is repeated 20 million times, and we report the Type I error down to $10^{-5}$. 
Table \ref{p2_tab:typeIerr} shows that our GBJ p-value calculation is accurate and protects the correct size
for correlation structures seen in actual data. 
The p-value calculation for the omnibus test is similarly accurate at the most stringent significance levels, but it is
somewhat conservative at larger significance levels.

%%%%%%%%%%%%%%%%%%%%%%%%%%%%%%%%%%%%%%%%%%
%%%%%%%%%%%%%%%%%%%%%%%%%%%%%%%%%%%%%%%%%%

\begin{table}[ht]
\caption{Type I error of GBJ computed over 20 million simulations. The strong LD setting denoted by SLD refers to
eight SNPs from FGFR2 which are highly correlated.  The weak LD setting denoted by WLD refers to eight SNPs from FGFR2 which
demonstrate only a small amount of correlation with each other. The Chr5 setting refers to over 500 regions on chromosome 5 corresponding to known gene sizes.}
\small
\begin{center}
\begin{tabular}{rrrrrrr}
  \hline
Nominal $\alpha$ & GBJ, WLD & GBJ, SLD & GBJ, Chr5 & OMNI, WLD & OMNI, SLD & OMNI, Chr5 \\ 
  \hline
$1\cdot10^{-2}$ & $9.65\cdot10^{-3}$ & $8.50\cdot10^{-3}$ & $9.35\cdot10^{-3}$ & $7.22\cdot10^{-3}$ & $6.88\cdot10^{-3}$ & $6.74\cdot10^{-3}$ \\ 
  $1\cdot10^{-3}$ & $9.70\cdot10^{-4}$ & $9.21\cdot10^{-4}$ & $1.06\cdot10^{-3}$ & $6.95\cdot10^{-4}$ & $7.19\cdot10^{-4}$ & $6.90\cdot10^{-4}$ \\ 
  $1\cdot10^{-4}$ & $9.82\cdot10^{-5}$ & $9.90\cdot10^{-5}$ & $1.20\cdot10^{-4}$ & $7.70\cdot10^{-5}$ & $9.00\cdot10^{-5}$ & $8.06\cdot10^{-5}$ \\ 
  $1\cdot10^{-5}$ & $9.65\cdot10^{-6}$ & $1.11\cdot10^{-5}$ & $1.49\cdot10^{-5}$ & $9.35\cdot10^{-6}$ & $1.31\cdot10^{-5}$ & $1.08\cdot10^{-5}$ \\ 
   \hline
\end{tabular}
\end{center}
\label{p2_tab:typeIerr}
\end{table}

%%%%%%%%%%%%%%%%%%%%%%%%%%%%%%%%%%%%%%%%%%
%%%%%%%%%%%%%%%%%%%%%%%%%%%%%%%%%%%%%%%%%%

\subsection{Power of GBJ Under Varying Hypothetical Correlation Structures and Sparsity Settings}
\label{p2_ss:structured_power}

To study the power of GBJ, we first conduct simulations under a variety
of hypothetical correlation structures and sparsity settings. 
The performance of the GBJ is compared to the minimum p-value test, GHC, SKAT, 
and the omnibus test described in Section \ref{p2_ss:omnibus}. 
The MinP test p-value is calculated by casting MinP as a boundary-defining test 
with $b_{j}=|Z|_{(d)}$ for all $j$, and then computation proceeds through
the methods described in Section \ref{p2_ss:pvalue}.
GBJ and GHC similarly use the p-value calculation described above.
For SKAT, we run the corresponding R package.

To study how power is impacted by different correlation structures between the SNPs, 
we utilize block correlation structures which are slightly more complex than those 
used for the rejection region analysis in Section \ref{p2_sec:rej_region}. 
Specifically, consider a set of causal SNPs that are correlated amongst themselves 
with common pairwise correlation $\rho_{1}$.
All other SNPs are then non-causal, and we allow half of them to have an exchangeable
correlation structure with correlation $\rho_{3}$; the other half of the non-causal SNPs 
are completely independent of all other non-causal SNPs. 
Finally the pairwise correlation between a causal SNP and a non-causal SNP is set at $\rho_{2}$.
The three correlations $\rho_{1},\rho_{2},\rho_{3}$ will vary between 0 and 0.3.  
All SNPs are generated to have minor allele frequency of 0.3.

\begin{figure}
\begin{center}
\centerline{\includegraphics[scale=0.28, angle=270]{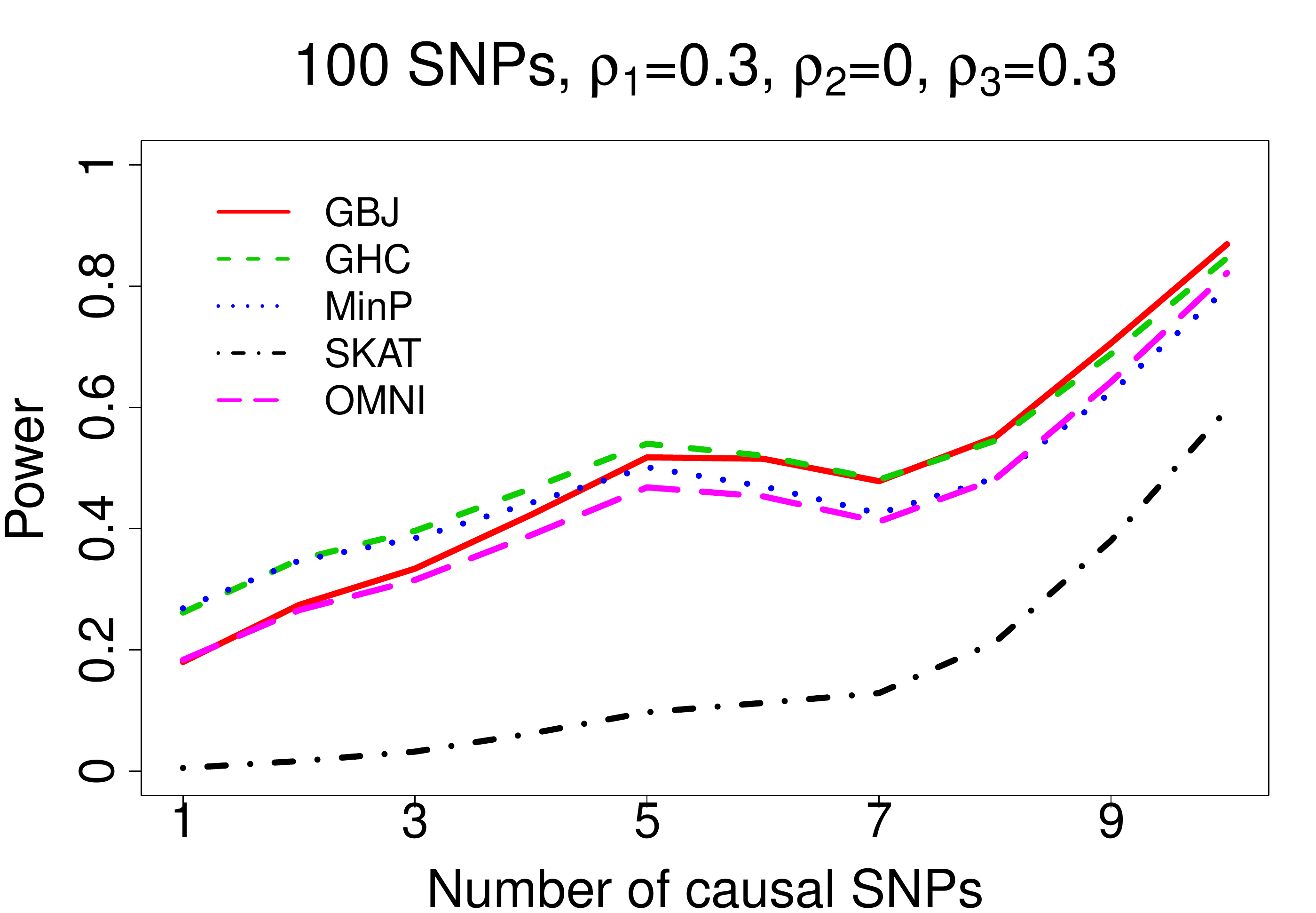}\includegraphics[scale=0.28, angle=270]{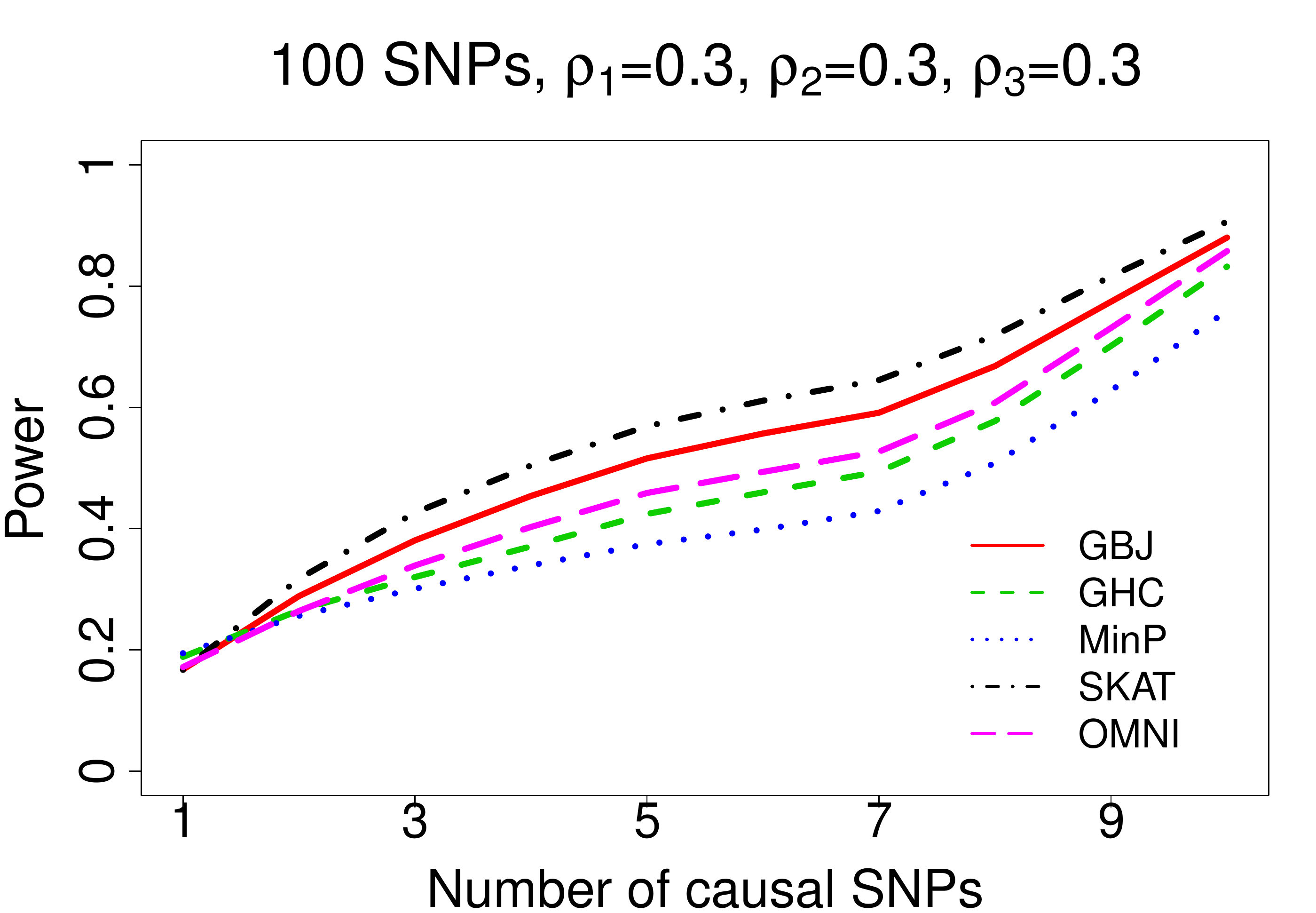}}
\end{center}
\caption{Power of set-based tests when there is correlation within noise SNPs (left) and across all SNPs (right).  
The correlation structure on the right is slightly simpler than the previous three structures, as we switch to an exchangeable correlation matrix in order to accommodate $\rho_{2}=0.3$ while still ensuring the correlation matrix is positive definite.  As the number of causal SNPs increases, we decrease the effect size to keep power bounded away from one.}
\label{p2_fig:pow_fig2}
\end{figure}

We demonstrate the effects of signal sparsity by using a large SNP-set of $d$=100 SNPs
and varying the number of causal SNPs from $k=1$ to $k=10$. 
This allows us to examine power profiles in the very sparse regime (one to three causal SNPs), 
in the moderately sparse regime (four to nine causal SNPs), and at the edge of the dense regime
(ten or more causal SNPs). 
The true disease model is 
\begin{equation}
Y_{i}=\sum_{j=1}^{k}\beta_{j}G_{ij}+\epsilon_{i},\epsilon_{i}\sim N(0,1),
\label{p2_eq:sim_model}
\end{equation}
where all the $\beta_{j}$ are the same and depend on the number of causal SNPs $k$.
We reduce $\beta_{j}$ slightly as $k$ increases in order to keep the power of each test 
below one throughout the entire sparse regime.
The full details on effect size for each simulation are available in the Supplementary Materials.
Figure \ref{p2_fig:pow_fig1} considers the case where the noise SNPs
are independent, and Figure \ref{p2_fig:pow_fig2} considers the case where the 
noise SNPs are correlated. 
We perform 500 simulations at each different value of $k$ and test
at $\alpha=0.01$.
All the power curves are smoothed to show empirical power.

%%%%%%%%%%%%%%%%%%%%%%%%%%%%%%%%%%%%%%%%
%%%%%%%%%%%%%%%%%%%%%%%%%%%%%%%%%%%%%%%%
\begin{figure}
\begin{center}
\centerline{\includegraphics[scale=0.28, angle=270]{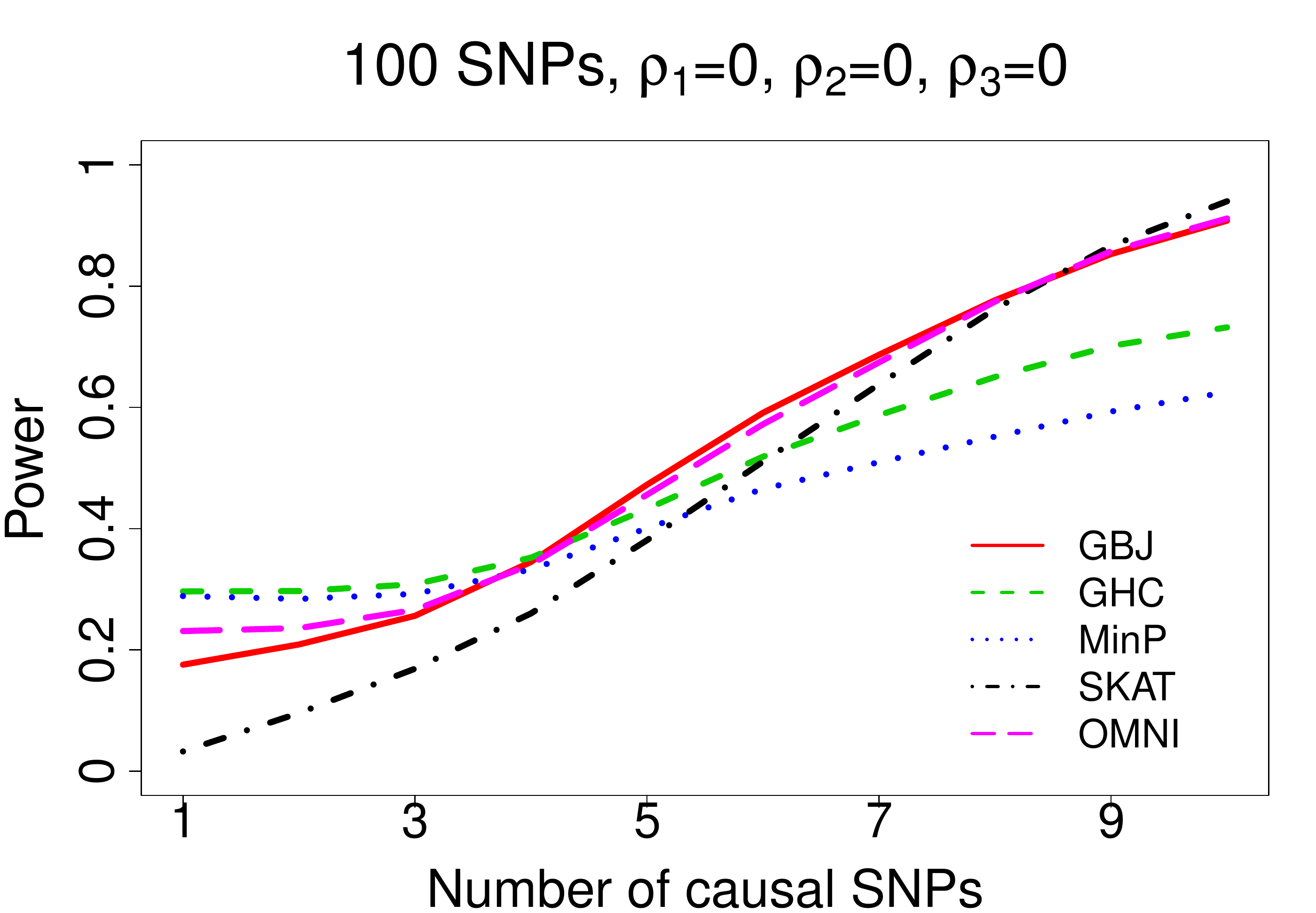}\includegraphics[scale=0.28, angle=270]{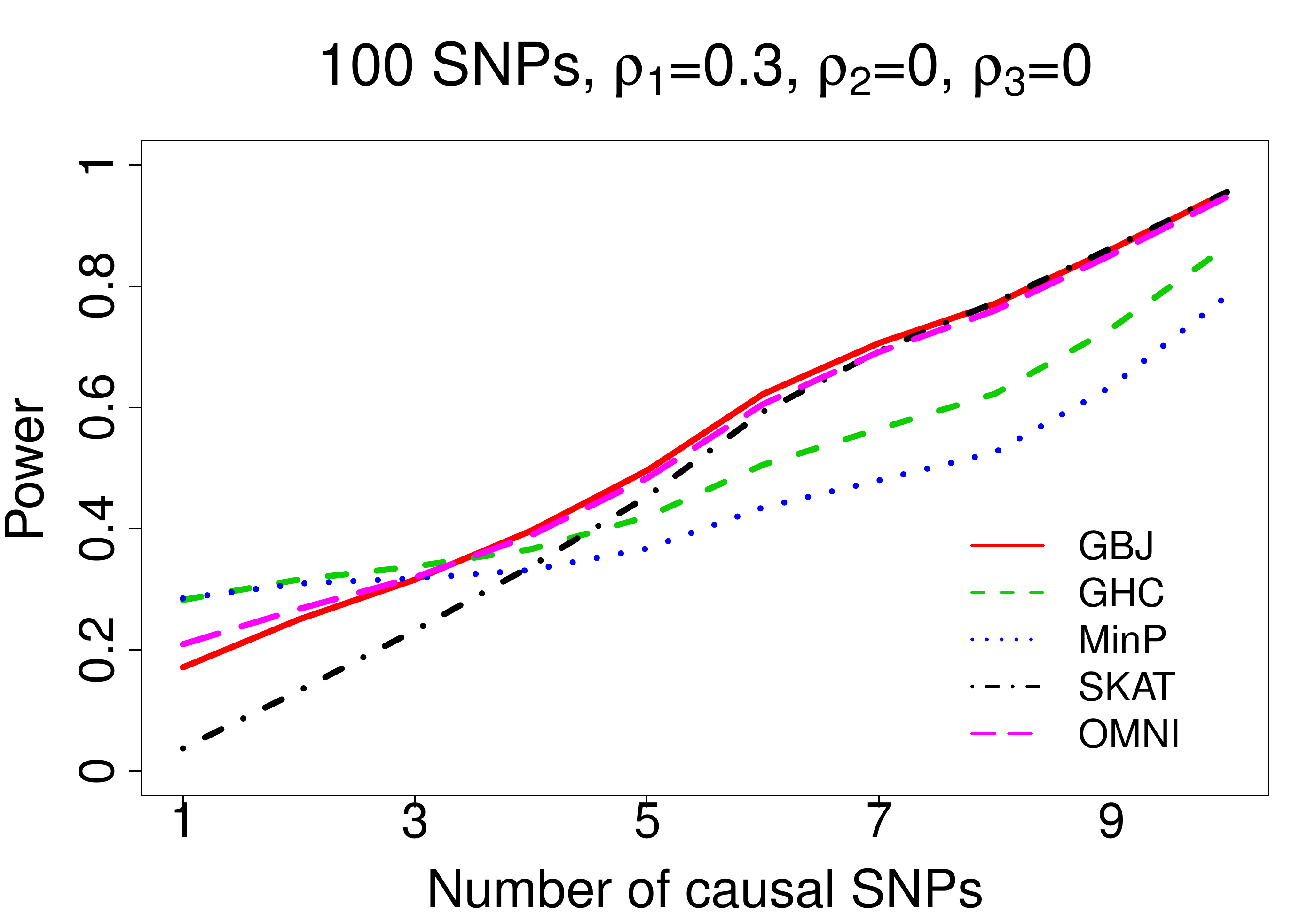}}
\end{center}
\caption{Power of set-based tests when noise SNPs are independent.  On the left, all SNPs are completely independent of each other. On the right, causal SNPs are correlated within themselves at $\rho_{1}=0.3$.  As the number of causal SNPs increases, we slightly decrease the effect size to keep power bounded away from one.}
\label{p2_fig:pow_fig1}
\end{figure}
%%%%%%%%%%%%%%%%%%%%%%%%%%%%%%%%%%%%%%%%
%%%%%%%%%%%%%%%%%%%%%%%%%%%%%%%%%%%%%%%%

%%%%%%%%%%%%%%%%%%%%%%%%%%%%%%%%%%%%%%%%
%%%%%%%%%%%%%%%%%%%%%%%%%%%%%%%%%%%%%%%%

%%%%%%%%%%%%%%%%%%%%%%%%%%%%%%%%%%%%%%%%
%%%%%%%%%%%%%%%%%%%%%%%%%%%%%%%%%%%%%%%%

The first significant trend appearing in Figure \ref{p2_fig:pow_fig1} is 
the effect of sparsity on power.  
Among the non-omnibus tests, we see that GHC and MinP perform well when the number of causal SNPs is low,
as these tests often have the most power in the very sparse regime.
In both panels of Figure \ref{p2_fig:pow_fig1}, the transition to GBJ having
more power than GHC and minP occurs in the moderately sparse regime. 
Then as the number of causal SNPs increases into the dense regime, 
SKAT begins to catch up and eventually becomes the most powerful test. 
This behavior matches our intuition as well as previously published simulation
results.
GHC and minimum p-value place excess weight on the most outlying observations, 
so they are well-tuned to detect the very sparse signals. 
The rejection region of GBJ is better-suited to find moderately sparse signals, and SKAT is known to perform well
with dense signals. 

The relationship between sparsity and power can be modified by the total amount 
of correlation.
In the left panel of Figure \ref{p2_fig:pow_fig2} we set $\rho_{1}=\rho_{3}=0.3$, which 
corresponds to the situation where causal SNPs and non-causal SNPs are correlated 
within themselves, but the two groups are independent of each other.
In this case, MinP and GHC become the top-performing tests for a larger range of 
sparsity settings, with GBJ losing some of its advantage in the moderately sparse regime.
SKAT has almost no power in these situations, as the signals are sparse and there is
no correlation between causal and non-causal SNPs.
It appears that a large amount of correlation between the non-causal SNPs is 
detrimental to the performance of GBJ. 
An explanation for this behavior can be found in the rejection region analysis of 
Figure \ref{p2_fig:rej_region_fig}. 
We see that the bounds of GBJ appear less favorable compared to GHC when the 
amount of correlation is high. 
Since over half of the SNPs in Figure \ref{p2_fig:pow_fig2} are correlated with $\rho_{1}=\rho_{3}=0.3$,
these settings represent a much larger amount of total correlation than was present in 
Figure \ref{p2_fig:pow_fig1}. 

In the right panel of Figure \ref{p2_fig:pow_fig2} we investigate the setting of $\rho_{2}\neq0$ 
by using an exchangeable correlation structure, and SKAT dominates
as the most powerful test across almost all sparsity levels. 
Here we break slightly from the above framework by using exchangeable correlation
 to accommodate $\rho_{2}=0.3$ while still ensuring the correlation matrix of the SNPs is positive definite.  
GBJ is a close second to SKAT under most sparsity settings with these parameters.
SKAT is known to have good performance in the presence of LD between causal SNPs 
and noise SNPs, which makes signals appear to be dense. 
The increased density of signals also buoys the performance of GBJ compared to
GHC and minimum p-value, which perform the worst under exchangeable correlation.

Never losing too much power compared to the best test, the omnibus test appears to 
be robust to LD structure and sparsity.
This behavior is expected as our omnibus approach integrates information from tests which perform well 
across multiple sparsity settings.
Thus we would anticipate that OMNI is more resilient than any single test.

GBJ also demonstrates good power in a variety of situations, 
and its overall strength and robustness across the entire sparse
regime are attractive properties.
In particular, GBJ is the best-performing test when the 
level of sparsity is moderate and there is weak correlation among the noise SNPs. 
GBJ is disadvantaged against minP and GHC when signals are extremely sparse or there
is excess correlation among noise SNPs, but it outperforms these tests as signals become more dense.
In contrast, GBJ provides slightly less power than SKAT as signals reach the dense regime or
when there is moderate correlation between causal and non-causal SNPs, but GBJ
is also much more robust than SKAT when signals are sparse and not correlated with 
noise SNPs.
Similar to OMNI, GBJ does not fall far behind the best test in any given situation, 
which suggests that GBJ is a good choice to use when the signal sparsity is unknown.
Power for the standard BJ is not plotted in the interest of space, but it behaves like a dense test, similar to 
SKAT, under correlation.  
This behavior again matches Figure \ref{p2_fig:rej_region_fig}, which showed that BJ is more suited to detect 
dense signals as the amount of correlation increases.

\subsection{Power of GBJ under Actual Chromosome 5 Correlation Structures}
\label{p2_ss:chr5_power}

We conduct one final simulation to investigate the power of Generalized Berk-Jones 
under the unstructured LD patterns found in real GWAS data.  
In this simulation, we choose blocks of 40 SNPs at random locations on chromosome 5, and then genotype
data are generated using HAPGEN2.
We choose 40 to again approximately match the characteristics of FGFR2.
There are 2000 blocks chosen for each sparsity level, and 10 simulations are performed on each block,
for a total of 20000 at each number of causal SNPs.
Again, the effect size is decreased as the number of causal SNPs increases, with the
outcome still generated from equation (\ref{p2_eq:sim_model}).
Testing is performed at $\alpha=1 \cdot 10^{-5}$ to mimic a practical analysis.

\begin{figure}
\begin{center}
\centerline{\includegraphics[scale=0.28, angle=270]{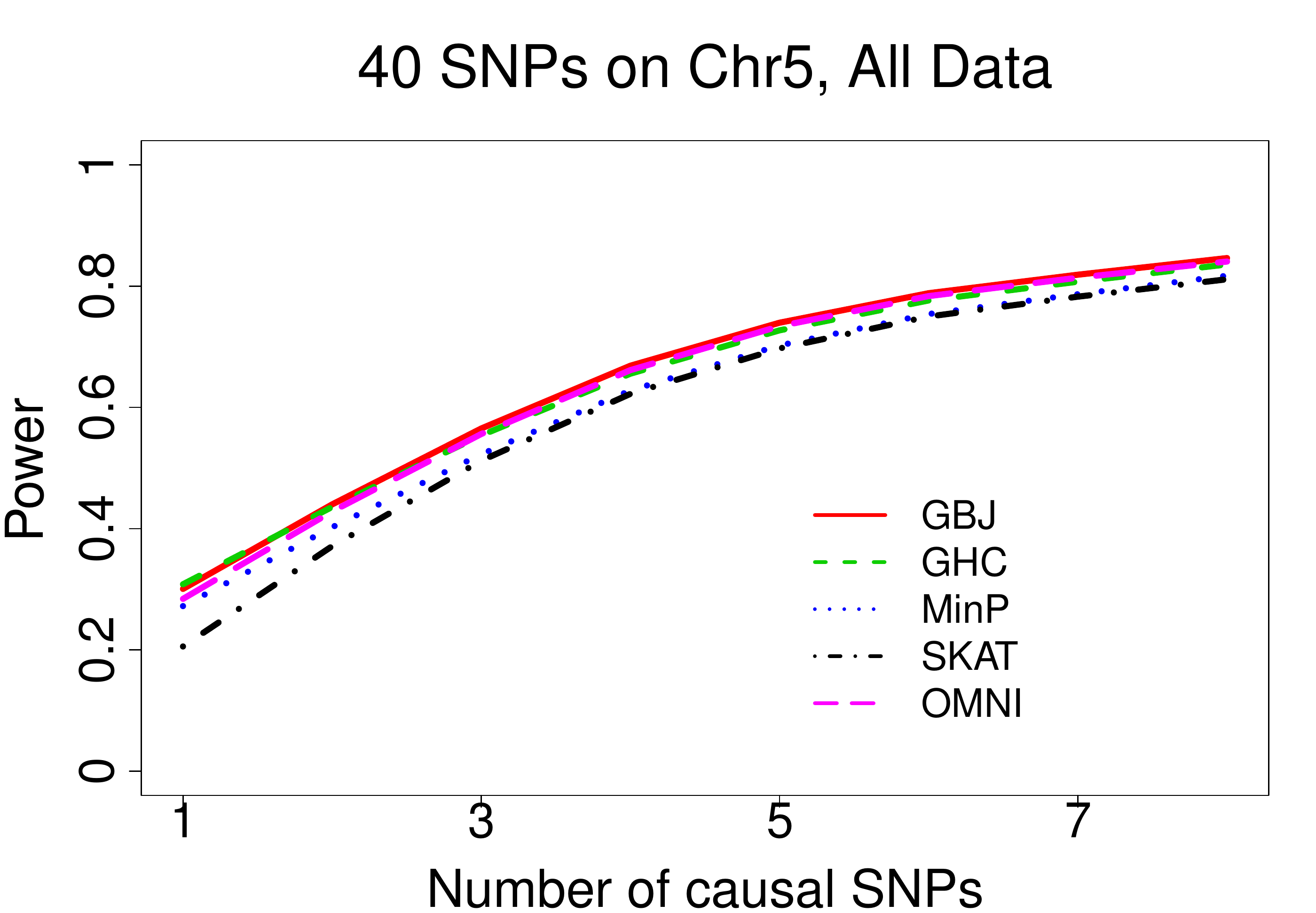}\includegraphics[scale=0.28, angle=270]{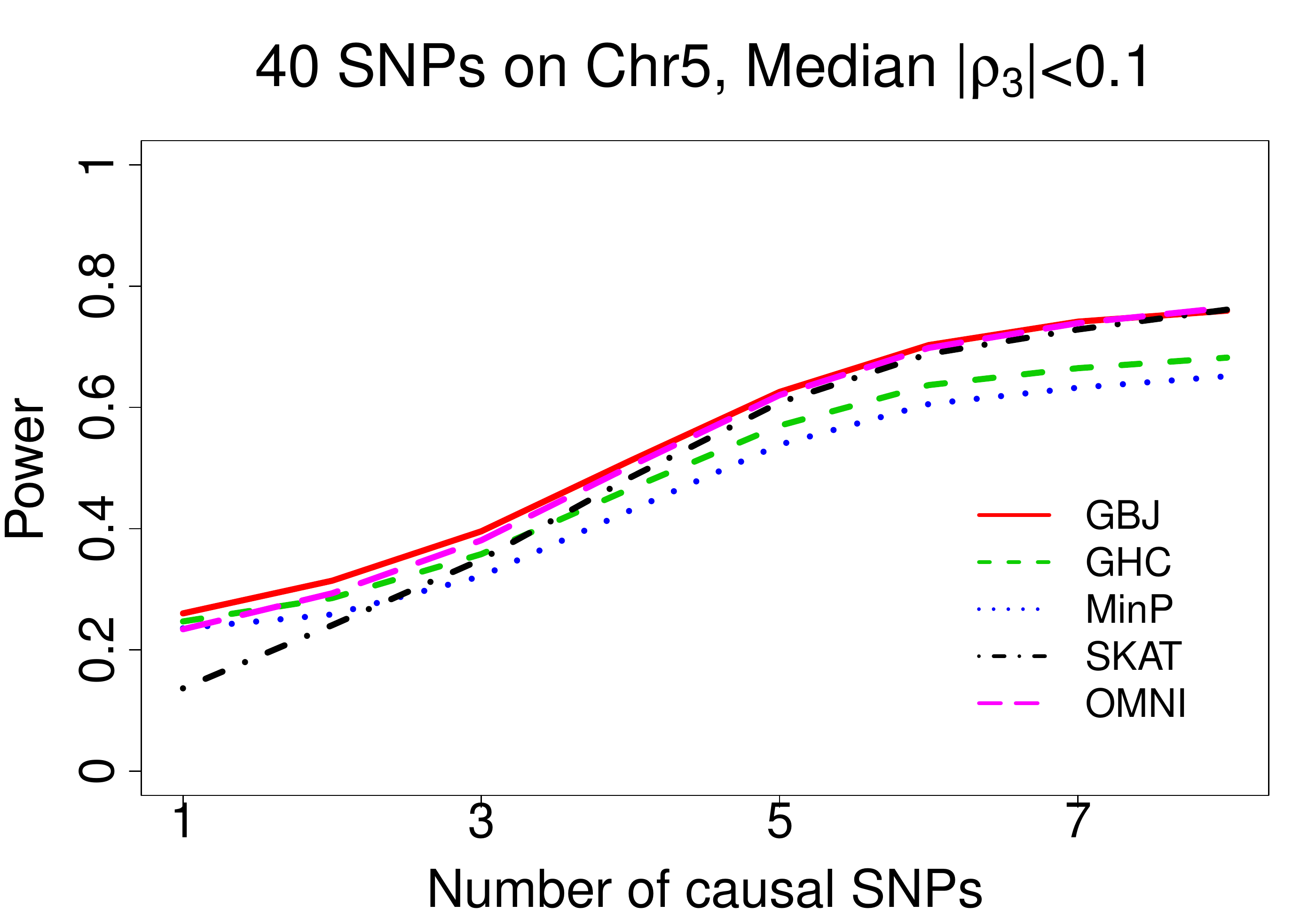}}
\end{center}
\caption{
Power of set-based tests with correlation structures found in actual chromosome 5 data.  Left panel presents all simulations and right panel shows only SNP-sets where median $|\rho_{3}|<0.1$. The effect size begins at $\beta_{j}=0.16$ and falls to $\beta_{j}=0.1$ so that power remains below one as the number of causal SNPs increases. GBJ is the best-performing test in both panels as the number of causal SNPs increases.}
\label{p2_fig:pow_fig_chr5}
\end{figure}

We see in the left panel of Figure \ref{p2_fig:pow_fig_chr5} that GBJ, GHC, and the omnibus test all
have very similar power curves in this setting, while SKAT and minP lag slightly behind.
As the number of causal SNPs increases, GBJ demonstrates the best power by a small amount.
These results are rather homogenous because sparsity levels are more coarse
and because the parameters are a mix of
the values defined in Figures \ref{p2_fig:pow_fig1} and \ref{p2_fig:pow_fig2}.

Recall that GBJ extended its advantage over GHC when there was less correlation among noise SNPs.
When restricting our analysis to the blocks which have median $|\rho_{3}| < 0.1$, we see
a more substantial superiority of GBJ over GHC in the moderately sparse regime, similar to Figure 
\ref{p2_fig:pow_fig1}.
Thus it is possible to recover the patterns of the structured simulation in real genotype data.
Obviously median $|\rho_{3}|$ is not a perfect summary measure, as it cannot single-handedly
capture all the parameters in a $40 \times 40$ correlation matrix.
Further parsing of the data would be necessary to see larger differences in performance.
In a practical setting, we might switch between tests based on certain SNP-set characteristics,
such as applying GBJ when the set is large and likely to have moderately sparse signals.
These results do again demonstrate the robustness of GBJ across multiple situations, as
it provides the most power across a large portion of the sparse regime.

%%%%%%%%%%%%%%%%%%%%%%%%%%%%%%%%%%%%%%%%
%%%%%%%%%%%%%%%%%%%%%%%%%%%%%%%%%%%%%%%%
%%%%%%%%%%%%%%%%%%%%%%%%%%%%%%%%%%%%%%%%
%%%%%%%%%%%%%%%%%%%%%%%%%%%%%%%%%%%%%%%%

%%%%%%%%%%%%%%%%%%%%%%%%%%%%%%%%%%%%%%%%%%%%%%%%%%%%%%%%%%%%
%%%%%%%%%%%%%%%%%%%%%%%%%%%%%%%%%%%%%%%%%%%%%%%%%%%%%%%%%%%%
%%%%%%%%%%%%%%%%%%%%%%%%%%%%%%%%%%%%%%%%%%%%%%%%%%%%%%%%%%%%

\section{Gene-Level Analysis of the CGEMS GWAS Data}
\label{p2_sec:CGEMS}

The CGEMS breast cancer dataset contains a case-control sample of
1145 breast cancer cases, all postmenopausal women
with European ancestry, and 1142 controls recruited from the Nurses' Health Study.
These women were genotyped at approximately 550000 SNPs with the Illumina HumanHap500 array.
The dataset was originally analyzed by \citet{CGEMS} in the single-marker GWAS approach.
The authors did not find any individual SNPs to reach the genome-wide significance level of $5\times 10^{-8}$,
but they highlighted FGFR2 as a strong candidate for future studies based on four SNPs in the gene 
that showed suggestive association with breast cancer.
Such a situation succinctly illustrates the burden of adjusting for multiple comparisons when 
testing individual SNPs.
Gene-level analysis provides an attractive alternative strategy that can reduce the number of 
comparisons and also aggregate evidence of signals across multiple SNPs in a gene. 
Here we perform a gene-level analysis to study the association between genes and breast cancer risk.

Since individual-level genotype data were available for this study, we first calculated the 
marginal test statistics for each SNP using the model in Section~\ref{p2_ss:IL_framework}. 
Specifically, we fit a logistic regression model with four covariates - age and the first three 
genotype principal components to control for population structure \citep{Eigenstrat}.
Then, for each of 14991 genes, we collected the marginal test statistics for all SNPs located 
within the region defined by that gene.
Each gene with more than one marginal SNP test statistic was analyzed with 
GBJ, GHC, SKAT, MinP, and the omnibus test.

In Table \ref{p2_tab:CGEMS_table}, we rank the top ten genes according to the smallest p-value produced by any of the five tests.
In this sample, GBJ provides the strongest evidence of association for the top four genes
and five of the top ten. 
Most of these genes are ranked highly by multiple other methods, however no other method provides the 
lowest p-value for more than two of the top ten genes. 
In fact, GHC and MinP produce the smallest p-value only once between the two of them. 
One possible explanation for the underperformance of GHC and MinP is that there may be 
multiple tagged SNPs surrounding the true causal loci for each of these genes, which could
create a lack of extremely sparse alternatives.

%%%%%%%%%%%%%%%%%%%%%%%%%%%%%%%%%%%%%%%%
%%%%%%%%%%%%%%%%%%%%%%%%%%%%%%%%%%%%%%%%
\begin{table}[ht]
\caption{Top significant genes in gene-level analysis of CGEMS breast cancer GWAS data, ranked by minimum p-value produced by any of the five tests. The test which produces the smallest p-value for each gene is highlighted in red. }
\begin{center}
\begin{tabular}{lrrrrrr}
   \hline
Gene & GHC & GBJ & MinP & SKAT & OMNI & $d$ \\ 
  \hline
FGFR2 & $2.84\cdot10^{-5}$ & \red{$4.58\cdot10^{-6}$} & $8.20\cdot10^{-5}$ & $3.32\cdot10^{-5}$ & $2.58\cdot10^{-5}$ & $35$ \\ 
  CNGA3 & $3.00\cdot10^{-4}$ & \red{$4.04\cdot10^{-5}$} & $1.75\cdot10^{-3}$ & $8.34\cdot10^{-5}$ & $1.84\cdot10^{-4}$ & $26$ \\ 
  PTCD3 & $1.21\cdot10^{-4}$ & \red{$5.50\cdot10^{-5}$} & $3.16\cdot10^{-4}$ & $1.87\cdot10^{-4}$ & $6.83\cdot10^{-5}$ & $12$ \\ 
  POLR1A & $9.58\cdot10^{-5}$ & \red{$6.19\cdot10^{-5}$} & $4.62\cdot10^{-4}$ & $4.23\cdot10^{-4}$ & $3.87\cdot10^{-4}$ & $17$ \\ 
  ZNF263 & $4.89\cdot10^{-4}$ & $3.90\cdot10^{-4}$ & $8.09\cdot10^{-4}$ & $1.26\cdot10^{-3}$ & \red{$6.84\cdot10^{-5}$} & $3$ \\ 
  VWA3B & $4.20\cdot10^{-4}$ & $2.32\cdot10^{-4}$ & $1.43\cdot10^{-3}$ & $1.48\cdot10^{-4}$ & $4.87\cdot10^{-4}$ & $51$ \\ 
  TBK1 & $7.04\cdot10^{-4}$ & $3.35\cdot10^{-4}$ & $1.27\cdot10^{-3}$ & \red{$1.48\cdot10^{-4}$} & $6.05\cdot10^{-4}$ & $11$ \\ 
  ABCA1 & $3.74\cdot10^{-3}$ & \red{$1.65\cdot10^{-4}$} & $7.92\cdot10^{-3}$ & $4.99\cdot10^{-4}$ & $2.26\cdot10^{-4}$ & $63$ \\ 
  MMRN1 & $2.31\cdot10^{-4}$ & $5.51\cdot10^{-4}$ & \red{$1.72\cdot10^{-4}$} & $3.34\cdot10^{-2}$ & $7.73\cdot10^{-4}$ & $10$ \\ 
  TIGD7 & $5.79\cdot10^{-4}$ & $3.78\cdot10^{-4}$ & $1.32\cdot10^{-3}$ & $1.33\cdot10^{-3}$ & \red{$2.05\cdot10^{-4}$} & $4$ \\ 
   \hline
   \end{tabular}
   \end{center}
\label{p2_tab:CGEMS_table}
\end{table}
%%%%%%%%%%%%%%%%%%%%%%%%%%%%%%%%%%%%%%%%
%%%%%%%%%%%%%%%%%%%%%%%%%%%%%%%%%%%%%%%%

The lowest p-value for any gene over all five tests is produced by testing FGFR2 with GBJ, 
supporting the conclusions of \citet{CGEMS}.  
Since FGFR2 appears to have signals coming from at least four different SNPs and contains 35 SNPs in total, 
it would seem to fall into the category of moderate signal sparsity, where GBJ has good performance. 
Thus we may have expected beforehand that GBJ would be the most powerful test for this gene.
FGFR2 has been further validated as a breast cancer associated locus in multiple follow-up studies 
\citep{FGFR2_meyer, FGFR2_jie}.

Besides FGFR2, genes such as PTCD3 and POLR1A have also been implicated as risk loci in 
independent investigations \citep{PTCD3, dmGWAS}.  
The overlap of our findings with other studies and other statistics provides a level
of reassurance that GBJ performs well in identifying truly significant genes and not simply 
spurious associations.
Alternately, ABCA1 is an example of a gene that may not have received further scrutiny if we 
were not utilizing the GBJ test. 
ABCA1 expression has been linked with breast cancer risk \citep{ABCA1}, but MinP and GHC 
do not provide the same strength of evidence that GBJ does. 
It seems likely that there are more than a few signal SNPs in ABCA1, especially since ABCA1 
contains a relatively large number of SNPs compared to the other genes in this dataset.

Perhaps due to the limited sample size, no test produces a p-value low enough to be 
declared significant after Bonferroni correction for 14991 genes.  
Still, this analysis highlights the advantages of Generalized Berk-Jones compared to alternative tests.  
The GBJ p-value for FGFR2 does come very close to the Bonferroni-corrected level 
($3.34 \times 10^{-6}$), and it certainly provides more evidence of association than the single SNP statistics.  
Additionally, GBJ often gives the highest measure of significance, and never the lowest, 
in the genes displayed, demonstrating its robustness across different set sizes and LD patterns.

%%%%%%%%%%%%%%%%%%%%%%%%%%%%%%%%%%%%%%%%%%%%%%%%%%%%%%%%%%%%
%%%%%%%%%%%%%%%%%%%%%%%%%%%%%%%%%%%%%%%%%%%%%%%%%%%%%%%%%%%%
%%%%%%%%%%%%%%%%%%%%%%%%%%%%%%%%%%%%%%%%%%%%%%%%%%%%%%%%%%%%

\section{Discussion}
\label{p2_sec:discussion}

We have proposed the Generalized Berk-Jones statistic to test for association
between a SNP-set and an outcome. Our GBJ generalizes the standard
Berk-Jones by modifying the BJ statistic to directly account for the correlation between individual SNPs.
This modification results in a test that is more powerful when SNPs are in LD.
We also provide an analytic p-value calculation for GBJ and generalize it to a class
of supremum-based global tests, allowing valid inference for HC, GHC, BJ, and other 
methods when these procedures are applied as SNP-set tests using correlated marginal test statistics. 
Rejection region analysis demonstrates that GBJ can be described as
a compromise between Berk-Jones and Higher Criticism-type tests in terms of finite sample performance.

While our numerical analysis shows situations where GBJ does not set the lowest boundary at 
either $|Z|_{(d)}$ or  $|Z|_{(d/2)}$, GBJ generally comes very close to the lowest boundary at both locations, 
which affords it both robustness to signal sparsity and power to detect moderately sparse signals.
GHC and HC often set the lowest boundary around $|Z|_{(d)}$, but in return they concede a large amount of
volume past the first few most extreme observations, which lowers power in the moderately sparse regime.
BJ frequently sets the lowest boundary past the tail, but its tail boundary can be orders of magnitude larger than
that of GBJ, HC, and GHC.
Bounds in the expected signal regions must be viewed holistically, so slightly lower
bounds at a few locations are not necessarily desirable if the price is much higher bounds in other signal locations, as
in the case of BJ.
Thus GBJ offers good power to detect moderately sparse effects without losing too much 
power when single-SNP signals are extremely sparse.

Simulation results reinforce the conclusions we find from examining the rejection regions of 
GHC and GBJ. 
Additionally we see that the MinP test performance is quite good when signals are very sparse,
similar to GHC, but MinP does not perform as well as GHC when signals become more dense.
SKAT has a unique power profile, as it can be particularly powerful when signals 
are dense or there is correlation between causal and non-causal SNPs,
 but it is also not robust to different correlation structures and will often have very 
little power in sparse settings where there is no correlation between causal and non-causal SNPs.
The omnibus test offers robust power across different sparsity levels, 
and while it is rarely the best test, it also never has the worst power.
When applied to data from the CGEMS study, we see that GBJ often produces the most 
significant p-values, perhaps owing to its versatility across different parameter settings. 

In demonstrating that the BJ statistic can be adapted for increased robustness to correlation, 
we have also demonstrated that these types of boundary-defining algorithms can be modified 
to increase finite sample power under specific set-level parameters.
It would be of interest to develop different boundary-defining methods that offer more favorable 
rejection regions in narrow but well-defined settings.
For example, it may be possible to define tests which outperform GHC or minP over
finer partitions of the extremely sparse regime.

In a similar vein, it would be interesting to understand the boundary shapes for other previously 
proposed boundary algorithms \citep{Phi_divergences} in the class of Berk-Jones and Higher Criticism.
While many of these algorithms share the same asymptotic guarantees of BJ and HC, 
little is known about their comparative finite sample performance, especially when observations in a set are correlated.
These other methods might also have great value in specific settings as mentioned above.
It would additionally be very valuable to develop SNP-set tests that are optimal in certain senses
for arbitrary sparsity and correlation.

As genomic data collection techniques continue to evolve, it may be necessary to adapt the GBJ as well.  
In particular, the rise of whole genome sequencing and fine mapping studies is leading to the discovery of more SNPs
with extremely rare minor alleles.
Marginal test statistics generated from these SNPs are known to be non-Gaussian in finite samples, and thus
they will not have the distribution we assume for GBJ.  
GBJ will need to account for null distributions that are not standard normal before SNP-sets 
containing rare variants can be tested.

\bigskip
\begin{center}
{\large\bf SUPPLEMENTARY MATERIAL}
\end{center}

The supplementary materials provide the proof of Theorem 1 from Section \ref{p2_ss:calculate_GBJ},  
offer further details on how to calculate the exact p-value from equation (\ref{p2_eq:exact_pvalue}) in Section \ref{p2_ss:pvalue},
and list the exact simulation parameters from Section \ref{p2_sec:simulation}.

\bibliographystyle{apalike}
\bibliography{ryan_refs_first2}

\begin{thebibliography}{}

\bibitem[{1000 Genomes Project Consortium}, 2015]{1000_Genomes}
{1000 Genomes Project Consortium} (2015).
\newblock A global reference for human genetic variation.
\newblock {\em Nature}, 526(7571):68--74.

\bibitem[Barnett et~al., 2017]{GHC}
Barnett, I., Mukherjee, R., and Lin, X. (2017).
\newblock The generalized higher criticism for testing {SNP-set} effects in
  genetic association studies.
\newblock {\em Journal of the American Statistical Association},
  112(517):64--76.

\bibitem[Berk and Jones, 1979]{BJ}
Berk, R.~H. and Jones, D.~H. (1979).
\newblock Goodness-of-fit test statistics that dominate the kolmogorov
  statistics.
\newblock {\em Probability Theory and Related Fields}, 47(1):47--59.

\bibitem[Boehm et~al., 2007]{PTCD3}
Boehm, J.~S., Zhao, J.~J., Yao, J., Kim, S.~Y., Firestein, R., Dunn, I.~F.,
  Sjostrom, S.~K., Garraway, L., Weremowicz, S., and Richardson, A. (2007).
\newblock Integrative genomic approaches identify {IKBKE} as a breast cancer
  oncogene.
\newblock {\em Cell}, 129:1065.

\bibitem[Conneely and Boehnke, 2007]{minP}
Conneely, K. and Boehnke, M. (2007).
\newblock So many correlated tests, so little time! {Rapid} adjustment of
  p-values for multiple correlated tests.
\newblock {\em The American Journal of Human Genetics}, 81:1158.

\bibitem[Dawson et~al., 2002]{Dawson_LD}
Dawson, E., Abecasis, G.~R., Bumpstead, S., Chen, Y., Hunt, S., Beare, D.~M.,
  Pabial, J., Dibling, T., Tinsley, E., Kirby, S., Carter, D., Papaspyridonos,
  M., Livingstone, S., Ganske, R., Lõhmussaar, E., Zernant, J., Tõonisson,
  N., Remm, M., Mägi, R., Puurand, T., Vilo, J., Kurg, A., Rice, K., Deloukas,
  P., Mott, R., Metspalu, A., Bentley, D.~R., Cardon, L.~R., and Dunham, I.
  (2002).
\newblock A first-generation linkage disequilibrium map of human chromosome 22.
\newblock {\em Nature}, 418(6897):544--548.

\bibitem[Donoho and Jin, 2004]{HC}
Donoho, D. and Jin, J. (2004).
\newblock Higher criticism for detecting sparse heterogeneous mixtures.
\newblock {\em Annals of Statistics}, 32(3):962--994.

\bibitem[Hunter et~al., 2007]{CGEMS}
Hunter, D., Kraft, P., Jacobs, K., Cox, D., Yeager, M., Hankinson, S.,
  Wacholder, S., Wang, Z., Welch, R., Hutchinson, A., and Wang, J. (2007).
\newblock A genome-wide association study identifies alleles in {FGFR2}
  associated with risk of sporadic postmenopausal breast cancer.
\newblock {\em Nature Genetics}, 39(7):870--874.

\bibitem[Jager and Wellner, 2007]{Phi_divergences}
Jager, L. and Wellner, J.~A. (2007).
\newblock Goodness-of-fit tests via phi-divergences.
\newblock {\em The Annals of Statistics}, 35(5):2018--2053.

\bibitem[Jia et~al., 2011]{dmGWAS}
Jia, P., Zheng, S., Long, J., Zheng, W., and Zhao, Z. (2011).
\newblock {dmGWAS}: dense module searching for genome-wide association studies
  in protein\-protein interaction networks.
\newblock {\em Bioinformatics}, 27(1):95--102.

\bibitem[Lee et~al., 2014]{rare_variant_review}
Lee, S., Abecasis, G.~R., Boehnke, M., and Lin, X. (2014).
\newblock Rare-variant association analysis: study designs and statistical
  tests.
\newblock {\em The American Journal of Human Genetics}, 95(1):5--23.

\bibitem[Li and Leal, 2008]{LiLeal}
Li, B. and Leal, S.~M. (2008).
\newblock Methods for detecting associations with rare variants for common
  diseases: application to analysis of sequence data.
\newblock {\em The American Journal of Human Genetics}, 83:311.

\bibitem[Li and Siegmund, 2015]{LiSiegmund}
Li, J. and Siegmund, D. (2015).
\newblock Higher criticism: $ p $-values and criticism.
\newblock {\em Annals of Statistics}, 43(3):1323--1350.

\bibitem[Liang et~al., 2008]{FGFR2_jie}
Liang, J., Chen, P., Hu, Z., Zhou, X., Chen, L., Li, M., Wang, Y., Tang, J.,
  Wang, H., and Shen, H. (2008).
\newblock Genetic variants in fibroblast growth factor receptor 2 ({FGFR2})
  contribute to susceptibility of breast cancer in chinese women.
\newblock {\em Carcinogenesis}, 29(12):2341--2346.

\bibitem[Manolio et~al., 2009]{Manolio_missing_herit}
Manolio, T.~A., Collins, F.~S., Cox, N.~J., Goldstein, D.~B., Hindorff, L.~A.,
  Hunter, D.~J., and McCarthy, M.~I. (2009).
\newblock Finding the missing heritability of complex diseases.
\newblock {\em Nature}, 461(7265):747--753.

\bibitem[McCullagh and Nelder, 1989]{GLM}
McCullagh, P. and Nelder, J.~A. (1989).
\newblock {\em Generalized Linear Models}.
\newblock CRC press.

\bibitem[Meyer et~al., 2008]{FGFR2_meyer}
Meyer, K.~B., Maia, A.-T., O'Reilly, M., Teschendorff, A.~E., Chin, S.-F.,
  Caldas, C., and Ponder, B.~A. (2008).
\newblock Allele-specific up-regulation of {FGFR2} increases susceptibility to
  breast cancer.
\newblock {\em PLoS Biology}, 1(5):e108.

\bibitem[Moscovich-Eiger and Nadler, 2017]{boundary_crossing}
Moscovich-Eiger, A. and Nadler, B. (2017).
\newblock Fast calculation of boundary crossing probabilities for poisson
  processes.
\newblock {\em Statistics \& Probability Letters}, 123:177--182.

\bibitem[Pasaniuc and Price, 2016]{summary_statistic_review}
Pasaniuc, B. and Price, A.~L. (2016).
\newblock Dissecting the genetics of complex traits using summary association
  statistics.
\newblock {\em Nature Genetics Reviews}, 18:117--127.

\bibitem[Prentice, 1986]{EBB}
Prentice, R.~L. (1986).
\newblock Binary regression using an extended beta-binomial distribution, with
  discussion of correlation induced by covariate measurement errors.
\newblock {\em Journal of the American Statistical Association},
  81(394):321--327.

\bibitem[Price et~al., 2006]{Eigenstrat}
Price, A.~L., Patterson, N.~J., Plenge, R.~M., Weinblatt, M.~E., Shadick,
  N.~A., and Reich, D. (2006).
\newblock Principal components analysis corrects for stratification in
  genome-wide association studies.
\newblock {\em Nature Genetics}, 38(8):904--909.

\bibitem[Smith and Land, 2012]{ABCA1}
Smith, B. and Land, H. (2012).
\newblock Anticancer activity of the cholesterol exporter {ABCA1} gene.
\newblock {\em Cell Reports}, 2.3:580--590.

\bibitem[Su et~al., 2011]{HAPGEN2}
Su, Z., Marchini, J., and Donnelly, P. (2011).
\newblock {HAPGEN2}: simulation of multiple disease snps.
\newblock {\em Bioinformatics}, 27(16):2304--2305.

\bibitem[Visscher et~al., 2012]{Visscher_five}
Visscher, P.~M., Brown, M.~A., McCarthy, M.~I., and Yang, J. (2012).
\newblock Five years of {GWAS} discovery.
\newblock {\em The American Journal of Human Genetics}, 90(1):7--24.

\bibitem[Walther, 2013]{ALR}
Walther, G. (2013).
\newblock The average likelihood ratio for large-scale multiple testing and
  detecting sparse mixtures.
\newblock In {\em From Probability to Statistics and Back: High-Dimensional
  Models and Processes}, volume~9, pages 317--326, Beachwood, OH. IMS.

\bibitem[Wu et~al., 2010]{snp_set}
Wu, M.~C., Kraft, P., Epstein, M.~P., Taylor, D.~M., Chanock, S.~J., Hunter,
  D.~J., and Lin, X. (2010).
\newblock Powerful {SNP-set} analysis for case-control genome-wide association
  studies.
\newblock {\em The American Journal of Human Genetics}, 86(6):929--942.

\bibitem[Wu et~al., 2011]{SKAT}
Wu, M.~C., Lee, S., Cai, T., Li, Y., Boehnke, M., and Lin, X. (2011).
\newblock Rare-variant association testing for sequencing data with the
  sequence kernel association test.
\newblock {\em The American Journal of Human Genetics}, 89(1):82--93.

\end{thebibliography}

\end{document}